\documentclass{article}
\usepackage{amssymb,latexsym, amsmath}
\newcommand{\bfi}{\bfseries\itshape}

\makeatletter
\@addtoreset{figure}{section}
\def\thefigure{\thesection.\@arabic\c@figure}
\def\fps@figure{h, t}
\@addtoreset{table}{bsection}
\def\thetable{\thesection.\@arabic\c@table}
\def\fps@table{h, t}
\@addtoreset{equation}{section}

\makeatother

\newtheorem{thm}{Theorem}[section]

\newtheorem{cor}[thm]{Corollary}

\begin{document}

\title{\vspace{-.5in}
The Euler--Poincar\'{e} Equations\\
           in Geophysical Fluid Dynamics}

\author{Darryl D. Holm
\\Theoretical Division and Center for Nonlinear Studies
\\Los Alamos National
Laboratory, MS B284
\\ Los Alamos, NM 87545
\\ {\footnotesize dholm@lanl.gov}
\and
Jerrold E. Marsden\\Control and Dynamical
Systems\\ California Institute of Technology 107-81\\ Pasadena, CA
91125
\\ {\footnotesize  marsden@cds.caltech.edu}
\and
Tudor S. Ratiu
\\Department of Mathematics
\\University of California, Santa Cruz, CA 95064
\\ {\footnotesize  ratiu@math.ucsc.edu}
\\ {\footnotesize To appear in the proceedings volume of the Isaac
Newton Institute for Mathematical Sciences}
\\ {\footnotesize {\it The Mathematics of Atmosphere and Ocean
Dynamics}}}
\date{April 16, 1998}

\maketitle

\begin{abstract}
Recent theoretical work has developed the Hamilton's-principle analog
of Lie-Poisson Hamiltonian systems defined on semidirect products.
The main theoretical results are twofold:
\begin{enumerate}
\item Euler--Poincar\'e equations (the Lagrangian analog of
Lie-Poisson Hamiltonian equations) are derived for a parameter
dependent Lagrangian from a general variational principle of Lagrange
d'Alembert type in which variations are constrained;

\item an abstract Kelvin--Noether theorem is derived for such systems.
\end{enumerate}
By imposing suitable constraints on the variations and by
using invariance properties of the Lagrangian, as one does for the
Euler equations for the rigid body and ideal fluids, we cast several
standard Eulerian models of geophysical fluid dynamics (GFD) at
various levels of approximation into Euler-Poincar\'{e} form and
discuss their corresponding Kelvin--Noether theorems and potential
vorticity conservation laws. The various levels of GFD approximation
are related by substituting a sequence of velocity decompositions and
asymptotic expansions into Hamilton's principle for the Euler
equations of a rotating stratified ideal incompressible fluid. We
emphasize that the shared properties of this sequence of approximate
ideal GFD models follow directly from their Euler-Poincar\'{e}
formulations. New modifications of the Euler-Boussinesq equations
and primitive equations are also proposed in which nonlinear
dispersion adaptively filters high wavenumbers and thereby enhances
stability and regularity without compromising either low wavenumber
behavior or geophysical balances.

\vspace{0.25in}
\footnoterule
\noindent
{\it Isaac Newton Institute for Mathematical Sciences, Proceedings,
1998}

\end{abstract}
\tableofcontents

\section{Introduction} \label{sec-intro}

The Eulerian formulation of the action principle for an ideal fluid
casts it into a form that is amenable to asymptotic expansions and
thereby facilitates the creation of approximate theories. This
Eulerian action principle is part of the general procedure of the
reduction theory of Lagrangian systems, including the theory
of the Euler--Poincar\'e equations (the Lagrangian analog of
Lie-Poisson Hamiltonian equations).  This setting provides a shared
structure for many problems in GFD, with several benefits, both
immediate (such as a systematic approach to hierarchical modeling and
versions of Kelvin's theorem for these models) and longer term (e.g.,
structured multisymplectic integration algorithms).

This paper will be concerned with Euler--Poincar\'e equations
arising from a family of action principles for a sequence of
standard GFD models in purely Eulerian variables at various levels
of approximation. We use the method of {\it Hamilton's principle
asymptotics} in this setting. In particular, the action principles of
these models are related by different levels of truncation of
asymptotic expansions and velocity-pressure decompositions in
Hamilton's principle for the unapproximated Euler equations of
rotating stratified ideal incompressible fluid dynamics. This
sequence of GFD models includes the Euler equations themselves,
followed by their approximations, namely: Euler-Boussinesq
equations (EB), primitive equations (PE), Hamiltonian balance
equations (HBE), and generalized Lagrangian mean (GLM) equations. We
also relate our approach to the rotating shallow water equations
(RSW), semigeostrophic equations (SG), and quasigeostrophic
equations (QG). Thus, asymptotic expansions and velocity-pressure
decompositions of Hamilton's principle for the Euler equations
describing the motion of a rotating stratified ideal incompressible
fluid will be used to cast the standard EB, PE, HBE and GLM models
of GFD into Euler-Poincar\'{e} form and thereby unify these
descriptions and their properties at various levels of approximation.
See Tables 4.1 and 4.2 for summaries.

These GFD models have a long history dating back at least to Rossby
[1940], Charney [1948] and Eliassen [1949], who used them, in their
simplest forms (particularly the quasigeostrophic and semigeostrophic
approximations), to study structure formation on oceanic and
atmospheric mesoscales. The history of the efforts to establish the
proper equations for synoptic motions is summarized by Pedlosky
[1987] and Cushman-Roisin [1994]; see also Phillips [1963]. One
may consult, for example, Salmon [1983, 1985, 1988], Holm, Marsden,
Ratiu and Weinstein [1985], Abarbanel, Holm, Marsden, and Ratiu
[1986], and Holm [1996] for recent applications of the approach of
Hamilton's principle asymptotics to derive approximate equations in
GFD.

Well before Rossby, Charney, and Eliassen, at the end of the 19th
century, Poincar\'e [1901] investigated the formulation of the
Euler equations for the dynamics of a rigid body in Lie algebraic
form. Poincar\'e's formulation of the Euler equations for a rigid
body carries over naturally to the dynamics of ideal continua, as
shown by Holm, Marsden and Ratiu [1998a], and Poincar\'e's ideas will
form the basis of the present study. Some of Poincar\'e's other key
papers in this area are listed in the bibliography.

Starting from the action principle for the Euler equations, the
present work first expresses the various GFD equations in the
Euler-Poincar\'{e} form for continua due to Holm, Marsden and Ratiu
[1998a] and discusses the properties acquired by casting the GFD
equations into this form. The main property so obtained is the
Kelvin--Noether theorem for the theory. This, in turn, leads to
conservation of potential vorticity on fluid parcels.
Domain-integrated energy is also conserved and the relation of the
Euler--Poincar\'e equations to the Lie-Poisson Hamiltonian
formulation of the dynamics is given by a Legendre transformation at
the level of the Lie algebra of divergenceless vector fields.

The methods of this paper are based on reduction of variational
principles; that is, on Lagrangian reduction (see Cendra et al.
[1986, 1987] and Marsden and Scheurle [1993a,b]), which is also
useful for systems with nonholonomic constraints. This has been
demonstrated in the work of Bloch, Krishnaprasad, Marsden and Murray
[1996], who derived the reduced La\-gran\-ge d'Al\-em\-bert
equations for such nonholonomic systems. Coupled with the methods of
the present paper, these techniques for handling nonholonomic
constraints should also be useful for continuum systems. In
addition, it seems likely that the techniques of multisymplectic
geometry, associated variational integrators, and the
multisymplectic reduction will be exciting developments for the
present setting; see Marsden, Patrick and Shkoller [1997] for the
beginnings of such a theory.

\paragraph{Organization of the Paper.} In \S\ref{sec-EP} we recall
from Holm, Marsden and Ratiu [1998a] the abstract Euler-Poincar\'{e}
theorem for Lagrangians depending on parameters along with the
associated Kelvin--Noether theorem. These theorems play a key role in
the rest of our analysis. In \S\ref{sec-EPPC} we discuss their
implications for continuum mechanics and then in
\S\ref{sec-GFD-applic} we apply them to a sequence of models in
geophysical fluid dynamics. We begin in \S\ref{sec-Euler} and
\S\ref{sec-EB} by recalling the action principles in the Eulerian
description for the Euler equations and their Euler-Boussinesq
approximation, respectively. Then we show how these standard GFD
models satisfy the Euler-Poincar\'{e} theorem.
These sections also introduce the scaling
regime and small parameters we use in making asymptotic expansions
and velocity-pressure decompositions that are used in the remaining
sections. Next, \S\ref{sec-PE} introduces the hydrostatic
approximation into the Euler-Poincar\'{e} formulation of the
Euler-Boussinesq equations to yield the corresponding formulation of
the primitive equations.  Later sections cast further approximations
of the Euler-Boussinesq equations into the Euler-Poincar\'{e}
formulation, starting in \S\ref{sec-HBE} with the Hamiltonian balance
equations and proceeding to the generalized Lagrangian mean (GLM)
theory for wave, mean flow interaction (WMFI), due to Andrews and
McIntyre [1978a,b] in \S\ref{sec-glmeqs}. In \S\ref{sec-MEB} we use
the Euler--Poincar\'e theorem, including advected parameters, to
formulate a new model of ideal GFD called the EB$\alpha$ model that
includes {\it nonlinear dispersion} along with stratification and
rotation. The EB$\alpha$ equations modify the usual Euler-Boussinesq
equations by introducing a length scale, $\alpha$. The length scale
$\alpha$ is interpreted physically in the GLM setting as the amplitude
of the rapidly fluctuating component of the flow. We derive the
Euler--Poincar\'e equations for the EB$\alpha$ model by making an
asymptotic expansion of the GLM Lagrangian for WMFI in powers of
$\alpha$ and the Rossby number. Thus, the EB$\alpha$ model is a WMFI
turbulence closure model for a rotating stratified incompressible
fluid. In this model, nonlinear dispersion (parameterized by $\alpha$)
acts to filter the high wavenumbers ($k>1/\alpha$) and thereby
enhances solution stability and regularity without compromising either
low wavenumber behavior ($k<1/\alpha$), or geophysical balances. We
also present the corresponding nonlinear dispersive modification of
the primitive equations, called the PE$\alpha$ model. The nonlinear
dispersive filtering of high wavenumber activity in the EB$\alpha$
and PE$\alpha$ models regularizes these equations and thereby makes
them good candidates for long term numerical integration.

\section{The Euler--Poincar\'e Equations, Semidirect\\
Products, and Kelvin's Theorem}\label{sec-EP}

Here we recall from Holm, Marsden and Ratiu [1998a] the statements of
the Euler--Poincar\'e equations and their associated Kelvin--Noether
theorem. In the next section, we will discuss these statements in
the context of continuum mechanics and then in the following section
apply them to a sequence of models in geophysical fluid dynamics.
Although there are several possible permutations of the conventions,
we shall state the Euler--Poincar\'e theorem for the case of
{\it right actions} and {\it right invariant Lagrangians}, which is
appropriate for fluids and, in particular, for the GFD situation.

\subsection{The Euler--Poincar\'e Equations and Semidirect Products}

\paragraph{Assumptions and Notation.} We shall begin with the
abstract framework which will be a convenient setting for the several
special cases of GFD to follow.
\begin{itemize}
\item Let $G$ be a Lie group and let $\mathfrak{g}$ be its Lie
algebra. We consider a vector space $V$ and assume we have a {\it
right\/} representation of $G$ on $V$. The group $G$ then acts in a
natural way on the {\it right\/} on the dual space $V^\ast$ (the
action by $g \in G$ on $V^\ast$ is the dual of the action by $ g
^{-1} $ on $V$). We denote the action of $g$ on an element $ v \in
V $ by $ v g $ and on an element $ a \in V^\ast $ by $ a g $. In
general we use this {\it concatenation notation} for group actions.
Then $G$ also acts by right translation on $ TG $ and hence it acts
on  $TG \times V^\ast$. We denote the action of a group element
$g$ on a point $(v_h, a)$ by $(v_g, a)g = (v_hg, ag)$.
\item Assume we have a Lagrangian $ L : T G \times V ^\ast
\rightarrow {\mathbb{R}}$ that is right $G$--invariant.
\item For each $a_0 \in V^\ast$, define the
Lagrangian $L_{a_0} : TG \rightarrow {\mathbb{R}}$ by
$L_{a_0}(v_g) = L(v_g, a_0)$. Then $L_{a_0}$ is right
invariant under the lift to $TG$ of the right action of
$G_{a_0}$ on $G$, where $G_{a_0}$ is the isotropy group of $a_0$
(that is, the subgroup of elements of $G$ that leave the element $ a
_0 \in V^\ast $ invariant).
\item  Right $G$--invariance of $L$ permits us to define the
{\bfi reduced Lagrangian} through the equation
$l: {\mathfrak g} \times V^\ast \rightarrow {\mathbb{R}}$ by
\[
l(v_gg^{-1}, ag^{-1}) = L(v_g, a).
\]
Conversely,  this relation defines for any
$l: {\mathfrak g} \times V^\ast \rightarrow
{\mathbb{R}} $ a right $G$--invariant function
$ L : T G \times V ^\ast
\rightarrow {\mathbb{R}} $.
\item For a curve $g(t) \in G, $ let
$\xi (t) := \dot{g}(t) g(t)^{-1}$ and define the curve
$a(t) = a_0g(t)^{-1}$, which is the unique solution of the
linear differential equation with time dependent coefficients $\dot
a(t) = -a(t)\xi(t)$ with initial condition $a(0) = a_0$.
\item Let $\mbox {\rm ad}_{\xi}: \mathfrak{g}
\rightarrow \mathfrak{g}$ be the {\it infinitesimal adjoint
operator}; that is, the linear map given by the Lie algebra bracket:
$\mbox {\rm ad}_{\xi}(\eta) = [
\xi, \eta ].$ Let
$\mbox {\rm ad}_{\xi}^{\ast} : \mathfrak{g}^\ast \rightarrow
\mathfrak{g}^\ast $ be the dual of the linear transformation
$\mbox {\rm ad}_{\xi}$.
\end{itemize}

\begin{thm}[Euler--Poincar\'e reduction.] \label{rarl}
The following are equivalent:
\begin{enumerate}
\item [{\bf i} ] Hamilton's variational principle
\begin{equation} \label{hamiltonprincipleright1}
\delta \int _{t_1} ^{t_2} L_{a_0}(g(t), \dot{g} (t)) dt = 0
\end{equation}
holds, for variations $\delta g(t)$
of $ g (t) $ vanishing at the endpoints.
\item [{\bf ii}  ] $g(t)$ satisfies the Euler--Lagrange
equations for $L_{a_0}$ on $G$.
\item [{\bf iii} ]  The constrained variational principle
\begin{equation} \label{variationalprincipleright1}
\delta \int _{t_1} ^{t_2}  l(\xi(t), a(t)) dt = 0
\end{equation}
holds on $\mathfrak g \times V ^\ast $, using variations of the form
\begin{equation} \label{variationsright1}
\delta \xi = \dot{\eta } - {\rm ad}_{\xi} \eta
= \dot{\eta } - [\xi , \eta ], \quad
\delta a =  -a\eta ,
\end{equation}
where $\eta(t) \in \mathfrak g$ vanishes at the
endpoints.
\item [{\bf iv}] The {\bfi Euler--Poincar\'{e} equations} hold on
$\mathfrak g \times V^\ast$
\begin{equation} \label{eulerpoincareright1}
\frac{d}{dt} \frac{\delta l}{\delta \xi} = -
\mbox {\rm ad}_{\xi}^{\ast} \frac{ \delta l }{ \delta \xi}
+ \frac{\delta l}{\delta a} \diamond a.
\end{equation}
\end{enumerate}
\end{thm}

We refer to Holm, Marsden and Ratiu [1998a] for the proof of this in
the abstract setting. We shall see some of the features of this result
in the concrete setting of continuum mechanics shortly.

\paragraph{Important Notation.} Following the notational conventions
of Holm, Marsden and Ratiu [1998a], we let $\rho _v: \mathfrak{g}
\rightarrow V$ be the linear map given by $\rho_v (\xi) = v \xi$
(the right action of $\xi$ on $ v \in V $), and let
$\rho _v^\ast: V ^\ast  \rightarrow \mathfrak{g}^{\ast}$ be its dual.
For $ a\in V^\ast$, we write
\[
  \rho _v^\ast a = v \diamond a  \in \mathfrak{g}^\ast \, ,
\]
which is a bilinear operation in $v$ and $a$. Continuing to use the
concatenation notation for Lie algebra actions, the $\mathfrak{g}
$--action on $\mathfrak{g} ^\ast$ and $V^\ast$ is defined to be {\it
minus} the dual map of the $\mathfrak{g} $--action on $\mathfrak{g} $
and $V$ respectively and is denoted by $\mu \xi$ and $a \xi$ for $\xi
\in\mathfrak{g} $, $\mu \in \mathfrak{g} ^\ast$, and $a\in V^\ast$.
The following is a useful way to write the definition of $v \diamond a
\in\mathfrak{g}^\ast$: for all $v \in V$, $a \in V ^\ast$ and $\xi\in
\mathfrak{g}$, we have (note minus sign)
\begin{equation}
\left \langle v \diamond a\,, \xi \right\rangle
= \left \langle a\,, v\xi \right\rangle
= -\,\left\langle a \xi, v\right \rangle\,.
\label{def-diamond}
\end{equation}

\paragraph{The Legendre Transformation.} As explained in Marsden and
Ratiu [1994], one normally thinks of passing from Euler--Poincar\'e
equations on a Lie algebra $\mathfrak{g}$ to Lie--Poisson equations
on the dual $\mathfrak{g}^\ast$ by means of the Legendre
transformation. In our case, we start with a Lagrangian on $
\mathfrak{g}\times V^\ast $ and perform a {\it partial} Legendre
transformation in the variable $\xi$ only, by writing
\begin{equation}\label{legendre}
\mu = \frac{\delta l}{\delta \xi}\,, \quad
h(\mu, a) = \langle \mu, \xi\rangle - l(\xi, a).
\end{equation}
Therefore, we have the formulae
\begin{equation}
\frac{\delta h}{\delta \mu} = \xi +\left \langle \mu, \frac{\delta
\xi}{\delta \mu} \right \rangle - \left \langle \frac{\delta
l}{\delta \xi}\,, \, \frac{\delta \xi}{\delta \mu} \right \rangle\,
= \,\xi
\quad\hbox{and}\quad
\frac{\delta h}{\delta a} = -\,\frac{\delta l}{\delta a}\,.
\end{equation}
One of the points is that, consistent with the examples, {\it we do not
attempt} to use the full Legendre transformation to make
Euler--Poincar\'e equations on $\mathfrak{g} \times V $ correspond to
Lie--Poisson equations on the dual space $\mathfrak{g}^\ast \times
V^\ast$. In fact, such attempts will fail because in most interesting
examples, the full Legendre transform will be degenerate (the heavy
top, compressible fluids, etc). It is for this reason that we take a
partial Legendre transformation. In this case, our Euler--Poincar\'e
equations on $\mathfrak{g} \times V^\ast$ will correspond to the
Lie--Poisson equations on $\mathfrak{g}^\ast \times V^\ast$. We next
briefly recall the Hamiltonian setting on $\mathfrak{g}^\ast \times
V^\ast$.

\paragraph{Lie--Poisson Systems on Semidirect Products.} Let $S =
G\,\circledS\,V$ be the semidirect product Lie group for right
actions. Explicitly, the conventions for
$S$ are the following: the multiplication has the expression
\begin{equation}
(g_1, v_1)(g_2, v_2) = (g_1g_2, v_2 + v_1g_2),
\end{equation}
the identity element is $(e, 0)$, and the inverse is given by
$(g, v)^{-1} = (g^{-1}, -vg^{-1})$. The Lie algebra
of $S$ is denoted
$\mathfrak{s} = \mathfrak g\,\circledS\,V$ and it has the bracket
operation given by
\begin{equation}
[(\xi_1, v_1),\,(\xi_2, v_2)] = ([\xi_1, \xi_2], \, v_1\xi_2 -
v_2\xi_1).
\end{equation}
Let $\mathfrak s^\ast$ denote the dual of $\mathfrak s$.
For a {\it right\/} representation of $G$ on $V$ the
($+$) Lie-Poisson bracket of two functions $f, k : \mathfrak s^\ast
\rightarrow {\mathbb{R}} $ has the expression
\begin{eqnarray}\label{rightLP}
\{f, k\} _+ (\mu, a) & = & \left\langle \mu,
\left[
\frac{\delta  f}{\delta \mu } , \frac{\delta k}{\delta \mu }
\right] \right\rangle
- \left\langle a, \frac{\delta k }{ \delta a}\frac{\delta f}{\delta
\mu } - \frac{\delta f}{\delta a}\frac{\delta k}{\delta \mu }\,
\right\rangle \, ,
\end{eqnarray}
where $\delta f / \delta \mu  \in \mathfrak{g}$, and
$\delta f / \delta a \in V$
are the functional derivatives of $f$.
Using the diamond notation (\ref{def-diamond}), the corresponding
Hamiltonian vector field for $h : \mathfrak s^\ast \rightarrow
{\mathbb{R}}$ is easily seen to have the expression
\begin{equation}\label{righttham}
X_h (\mu, a) = - \left( {\rm ad}^\ast _{\delta h /\delta \mu}\mu
+\frac{\delta h }{ \delta a} \diamond a,\,
a\,\frac{ \delta h}{\delta \mu}\right)\,.
\end{equation}
Thus, Hamilton's equations on the dual of a semidirect product are
given by
\begin{eqnarray}
\dot{ \mu } & = & \{\mu, h\}
= -\,  {\rm ad}^\ast _{\delta h /\delta \mu}\mu
  -  \frac{\delta h }{ \delta a} \diamond a\,,
\label{rightsemi1.eqn} \\
\dot{ a } & = & \{a, h\}
= -\, a\, \frac{ \delta h}{\delta \mu}\,.
\label{rightsemi2.eqn}
\end{eqnarray}
where overdot denotes time derivative. Thus, the partial
Legendre transformation (\ref{legendre}) maps the Euler--Poincar\'e
equations (\ref{eulerpoincareright1}), together with the equations
$\dot{a} = - a(t)\xi(t)$ for $a$ to the Lie--Poisson equations
(\ref{rightsemi1.eqn}) and (\ref{rightsemi2.eqn}).

\paragraph{Cautionary Remark.}  If the vector space $V$ is absent
and one has just the equations
\begin{equation} \label{basiceulerpoincareright}
\frac{d}{dt} \frac{\delta l}{\delta \xi} = -
\mbox {\rm ad}_{\xi}^{\ast} \frac{ \delta l }{ \delta \xi}
\end{equation}
for $\xi \in \mathfrak{g}$ on a Lie algebra, one speaks of them as
the {\bfi basic Euler--Poincar\'e equations}. As explained in Holm,
Marsden and Ratiu [1998a], the Euler--Poincar\'e equations
(\ref{eulerpoincareright1}) {\it are not} the basic
Euler--Poincar\'e equations on the larger semidirect product Lie
algebra $\mathfrak g\,\circledS\, V^\ast$. This is a critical
difference between the Lie--Poisson and the Euler--Poincar\'e cases.
\paragraph{Advected Parameters.} As we shall see in the examples, and
as indicated by the above Euler--Poincar\'e reduction theorem, the
parameters $a\in V^\ast$ acquire dynamical meaning under Lagrangian
reduction. For the heavy top, the parameter is the unit vector in
the direction of gravity, which becomes a dynamical variable in the
body representation. For stratified incompressible fluids, the
parameters are the buoyancy $b$ and volume $D$ of a fluid element in
the reference configuration, which in the spatial representation
become dynamical variables satisfying the passive scalar advection
equation and continuity equation, respectively.

\subsection{The Kelvin--Noether Theorem} In this section, we explain
a version of the Noether theorem that holds for solutions of the
Euler--Poincar\'e equations. Our formulation is motivated and
designed for continuum theories (and hence the name
Kelvin--Noether), but it may be also of interest for finite dimensional
mechanical systems. Of course it is well known that the Kelvin
circulation theorem for ideal flow is closely related to the Noether
theorem applied to continua using the particle relabelling symmetry
group (see, for example, Arnold [1966]).

\paragraph{The Kelvin--Noether Quantity.} We start with a Lagrangian
$L _{a _0}$ depending on a parameter $a _0 \in V ^\ast$ as above. We
introduce a manifold ${\mathcal C}$ on which $G$ acts (on the right,
as above) and suppose we have an equivariant map
$\mathcal{K} : {\mathcal C} \times V ^\ast
\rightarrow \mathfrak{g} ^{\ast \ast} $.

In the case of continuum theories, the space ${\mathcal C}$ will be a
loop space and $\left\langle \mathcal{K} (c, a), \mu\right\rangle$ for
$c \in {\mathcal C}$ and $\mu \in
\mathfrak{g}^\ast$ will be  a circulation. This class of examples
also shows why we {\it do not} want to identify the double dual
$\mathfrak{g} ^{\ast \ast}$ with $\mathfrak{g}$.

Define the {\bfi Kelvin--Noether quantity}
$I : {\mathcal C} \times \mathfrak{g} \times V ^\ast
\rightarrow {\mathbb{R}}$ by
\begin{equation}\label{KelvinNoether}
I(c, \xi, a) = \left\langle\mathcal{K} (c, a), \frac{ \delta
l}{\delta \xi}( \xi , a)
\right\rangle.
\end{equation}

\begin{thm}[Kelvin--Noether] \label{KelvinNoetherthm}Fixing $c_0 \in
{\mathcal C}$, let $\xi (t), a(t)$ satisfy the
Euler--Poincar\'e equations and define $g(t)$ to be the solution of
$\dot{g}(t) = \xi(t)g(t)$ and, say, $ g (0) = e$. Let
$c(t) = c_0 g(t)^{-1}$ and $I(t) = I(c(t), \xi(t), a(t))$. Then
\begin{equation}
\frac{d}{dt} I(t) = \left\langle \mathcal{K}(c(t), a (t) ),
                     \frac{\delta l}{\delta a} \diamond a
\right\rangle.
\end{equation}
\end{thm}

\noindent{\bf Proof.\,} First of all, write $ a (t) = a_0
g(t)^{-1}$ and use equivariance to write
$I(t)$ as follows:
\[
   \left\langle \mathcal{K} (c(t), a(t) ) ,
          \frac{\delta l}{\delta \xi} (\xi(t), a(t)) \right\rangle
   =  \left\langle \mathcal{K} ( c_0, a _0 ),
     \left[ \frac{\delta l}{\delta \xi} (\xi(t),a(t)) \right]g(t)
\right\rangle
\]
The $g ^{-1}$ pulls over to the right side as $g$ (and not $g^{-1}$)
because of our conventions of always using right representations. We
now differentiate the right hand side of this equation. To do so, we
use the following well known formula for differentiating the
coadjoint action (see Marsden and Ratiu [1994], \S9.3):
\[
\frac{d}{dt} [\mu (t) g(t)]
=  \left[ {\rm ad}_{\xi (t)}^\ast  \mu (t)
     +   \frac{d}{dt} \mu (t) \right] g(t),
\]
where $\mu\in\mathfrak{g}^\ast$, and $\xi\in\mathfrak{g}$ is given by
\[ \xi (t) = \dot{g} (t) g (t) ^{-1} .
\]
Using this and the
Euler--Poincar\'e equations, we get
\begin{eqnarray*}
      \frac{d}{dt} I
  & = &  \frac{d}{dt} \left\langle \mathcal{K} ( c_0, a_0 ),
     \left[ \frac{\delta l}{\delta \xi} (\xi(t),a(t)) \right] g(t)
\right\rangle \\
  & = & \left\langle \mathcal{K} ( c _0, a_0 ) , \frac{d}{dt}
\left\{\left[ \frac{\delta l}{\delta \xi} (\xi(t),a(t)) \right]
g(t) \right\} \right\rangle \\
  & = & \left\langle
      \mathcal{K} (c_0, a_0), \left[{\rm ad}^\ast _\xi
       \frac{\delta l}{\delta \xi}
      - {\rm ad} _\xi ^\ast \frac{\delta l}{\delta \xi}
      + \frac{\delta l}{\delta a} \diamond a \right] g(t)
\right\rangle \\
& = & \left\langle \mathcal{K} (c_0, a_0 ) , \left[
\frac{\delta l}{\delta a} \diamond a\right] g(t) \right\rangle \\
& = & \left\langle { \mathcal K} ( c _0, a_0 ) g(t) ^{-1} , \left[
\frac{\delta l}{\delta a} \diamond a\right] \right\rangle \\
& = & \left\langle \mathcal{K} ( c (t), a (t)  ) ,  \left[
\frac{\delta l}{\delta a} \diamond a\right] \right\rangle
\end{eqnarray*}
where, in the last steps we used the definitions of the coadjoint
action, the Euler--Poincar\'e equation (\ref{eulerpoincareright1})
and the equivariance of the map $ \mathcal{K} $.
\quad
$\blacksquare$
\medskip

Because the advected terms are absent for the basic Euler--Poincar\'e
equations, we obtain the following.

\begin{cor} For the basic Euler--Poincar\'e equations, the Kelvin
quantity $I(t)$, defined the same way as above but with
$I : {\mathcal C} \times \mathfrak{g} \rightarrow{\mathbb{R}}$, is
conserved.
\end{cor}

\section{The Euler--Poincar\'e Equations in Continuum Mechanics}
\label{sec-EPPC}

In this section we will apply the Euler--Poincar\'e equations to the
case of continuum mechanical systems.

\paragraph{Vector Fields and Densities.} Let $\mathcal{D}$ be a bounded
domain in $\mathbb{R}^n$ with smooth boundary $\partial
\mathcal{D}$ (or, more generally, a smooth compact manifold with
boundary and given volume form or density). We let
$\mbox {\rm Diff}(\mathcal{D})$ denote the diffeomorphism group of
$\mathcal{D}$ of an appropriate Sobolev class (for example, as in
Ebin and Marsden [1970]). If the domain $\mathcal{D}$ is not compact,
then various decay hypotheses at infinity need to be imposed. Under
such conditions, $\mbox {\rm Diff}(\mathcal{D})$ is a smooth infinite
dimensional manifold and a topological group relative to the induced
manifold  topology. Right translation is smooth but left translation
and inversion are only continuous. Thus,
$\mbox {\rm Diff}(\mathcal{D})$ is not literally a Lie group in the
naive sense and so the previous theory must be applied with care.
Nevertheless, if one uses right translations and right
representations, the Euler--Poincar\'e equations of Theorem
\ref{rarl} do make sense, as a direct verification shows. We shall
illustrate such computations, by verifying several key facts
in the proof as we proceed.

Let $\mathfrak{X}(\mathcal{D})$ denote the space of
vector fields on $\mathcal{D}$ of the same differentiability class as
$\mbox{\rm Diff}(\mathcal{D})$.
Formally, this is the {\it right\/} Lie algebra of $\mbox{\rm
Diff}(\mathcal{D})$, that is, its standard {\it left\/} Lie algebra
bracket is {\it minus\/} the usual Lie bracket for vector fields. To
distinguish between these brackets, we shall reserve in what follows
the notation $[\mathbf{u},\,\mathbf{w}]$ for the standard Jacobi-Lie
bracket of the vector fields
$\mathbf{u},\,\mathbf{w}\in \mathfrak{X}({\mathcal{D}})$ whereas the
notation $\mbox {\rm ad}_\mathbf{u} \mathbf{w} :=
-[\mathbf{u},\,\mathbf{w}]$ denotes the adjoint action of the {\it
left\/} Lie algebra on itself. (The sign conventions will also be
clear in the coordinate expressions.)

We also let $\mathfrak{X}(\mathcal{D})^\ast$ denote the geometric dual
space of $\mathfrak{X}(\mathcal{D})$, that is,
$\mathfrak{X}(\mathcal{D})^\ast := \Omega^1({\mathcal{D}}) \otimes {\rm
Den}({\mathcal{D}})$,
the space of one--form densities on $\mathcal{D}$. If $\alpha \otimes
m \in \Omega^1({\mathcal{D}}) \otimes {\rm Den}({\mathcal{D}})$, the
pairing of  $\alpha \otimes m$ with $\mathbf{w} \in
\mathfrak{X}(\mathcal{D})$ is given by
\begin{equation}\label{continuumpairing}
\langle \alpha \otimes m, \mathbf{w} \rangle
= \int_{\mathcal{D}} \alpha\cdot \mathbf{w}\, m
\end{equation}
where $\alpha\cdot \mathbf{w}$ denotes the contraction of a
one--form with a vector field. For $\mathbf{w} \in
\mathfrak{X}(\mathcal{D})\,$ and $\alpha\otimes m \in
\mathfrak{X}(\mathcal{D})^\ast$, the dual of the adjoint
representation is defined by (note the minus sign)
\[
 \langle \mbox {\rm ad}^\ast_\mathbf{w}(\alpha\otimes m), \mathbf{u}
\rangle
= -\int_{\mathcal{D}}\alpha \cdot [\mathbf{w}, \mathbf{u}]\;m
\]
and its expression is
\begin{equation}\label{continuumcoadjoint}
\mbox {\rm ad}^\ast_\mathbf{w}(\alpha\otimes m) = (\pounds_\mathbf{w}
\alpha + (\mbox {\rm div}_m\mathbf{w})\alpha)\otimes m
= \pounds_\mathbf{w}(\alpha\otimes m)\,,
\end{equation}
where  ${\rm div}_m\mathbf{w}$ is the divergence of $\mathbf{w}$
relative to the measure $m$,, which is related to the Lie derivative by
$\pounds_\mathbf{w}m = ({\rm div}_m\mathbf{w})m$. Hence, if
$\mathbf{w} = w^i
\partial/\partial x^i,$ and $\alpha  = \alpha_i dx^i$, the one--form
factor in the preceding formula for $\mbox {\rm
ad}^\ast_\mathbf{w}(\alpha\otimes m)$ has the coordinate expression
\[
 \left ( w^j \frac{\partial
\alpha_i}{\partial
x^j} + \alpha_j \frac{\partial w^j}{\partial x^i} +
(\mbox {\rm div}_m\mathbf{w})\alpha_i \right )dx^i\,= \,
\left (\frac{\partial}{\partial x^j}(w^j\alpha_i) +
\alpha_j \frac{\partial w^j}{\partial x^i}\right ) dx^i\;,
\]
the last equality assuming that the divergence is taken
relative to the standard measure
$m = d^n\mathbf{x}$ in $\mathbb{R}^n$.

\paragraph{Configurations, Motions and Material Velocities.} Throughout
the rest of the paper we shall use the following conventions and
terminology for the standard quantities in continuum mechanics.
Elements of
$\mathcal{D}$ representing the material particles of the system are
denoted by $X$; their coordinates $X^A, A=1,...,n$ may thus be
regarded as the particle labels. A {\bfi configuration}, which we
typically denote by
$\eta$, is an element of ${\rm Diff}(\mathcal{D})$. A {\bfi motion}
$\eta_t$ is a path in ${\rm Diff}(\mathcal{D})$. The {\bfi
Lagrangian} or {\bfi material velocity\/} $\mathbf{U} (X,t)$  of the
continuum along the motion $\eta_t$ is defined by taking the time
derivative of the motion keeping the particle labels (the reference
particles) $X$ fixed:
\[
\mathbf{U} (X, t) := \frac{d\eta_t(X)}{dt}:=
\left.\frac{\partial}{\partial t}\right|_{X}\eta_t(X),
\]
the second equality being a convenient shorthand notation of the
time derivative holding $X$ fixed.

Consistent with this definition of velocity, the tangent space to
${\rm Diff}(\mathcal{D})$ at $\eta \in
{\rm Diff}(\mathcal{D})$ is given by
\[
T_\eta {\rm Diff}(\mathcal{D})
= \{ \mathbf{U} _\eta: \mathcal{D} \rightarrow T {\mathcal{D}}
\mid \mathbf{U} _\eta(X)
\in T_{\eta(X)}\mathcal{D}\}.
\]
Elements of $T_\eta \mbox {\rm Diff}(\mathcal{D})$ are usually
thought of as vector fields on $\mathcal{D}$ covering $\eta$. The
tangent lift of right and left translations on $T\mbox {\rm
Diff}(\mathcal{D})$ by $\varphi \in \mbox {\rm Diff}(\mathcal{D})$
have the expressions
\[
\mathbf{U} _\eta\varphi := T_\eta R_\varphi (\mathbf{U} _\eta)
= \mathbf{U} _\eta \circ \varphi\,\qquad {\rm and} \qquad
\varphi\mathbf{U} _\eta := T_\eta L_\varphi (\mathbf{U} _\eta)
= T\varphi \circ \mathbf{U} _\eta \,.
\]

\paragraph{Eulerian Velocities.} During a motion
$\eta_t$, the particle labeled by $X$ describes a path in
$\mathcal{D}$ whose points $x(X, t):= \eta_t(X)$ are
called the {\bfi Eulerian} or {\bfi spatial points\/} of this path.
The derivative $\mathbf{u}(x, t)$ of this path, keeping the
Eulerian point $x$ fixed, is called the {\bfi Eulerian}
or {\bfi spatial velocity\/} of the system:
\[
\mathbf{u}(x, t):= \mathbf{U} (X, t) :=
\left.\frac{\partial}{\partial t}\right|_x\eta_t(X).
\]
Thus, the Eulerian velocity $\mathbf{u}$ is a time dependent vector
field on $\mathcal{D}$: $\mathbf{u}_t \in
\mathfrak{X}(\mathcal{D})$, where $\mathbf{u}_t(x) := \mathbf{u}(x,
t)$. We also have the fundamental relationship
\[
\mathbf{U} _t = \mathbf{u}_t \circ \eta_t\,,
\]
where $\mathbf{U} _t(X):= \mathbf{U} (X, t)$.
\medskip

The representation space $V^\ast$ of $\mbox {\rm Diff}(\mathcal{D})$
in continuum mechanics is often some subspace of $\mathfrak{T}
(\mathcal{D})\otimes {\rm Den}({\mathcal{D}})$, the tensor field
densities  on $\mathcal{D}$ and the representation is given by pull
back. It is thus a {\it right\/} representation of
$\mbox {\rm Diff}(\mathcal{D})$ on $\mathfrak{T}(\mathcal{D})\otimes
{\rm Den}({\mathcal{D}})$. The right action of the Lie algebra
$\mathfrak{X}({\mathcal{D}})$ on $V^\ast$ is given by $ a\mathbf{u}
:= \pounds_\mathbf{u} a$, the Lie derivative of the tensor field
density $a$ along the vector field
$\mathbf{u}$.

\paragraph{The Lagrangian.} The Lagrangian of a continuum mechanical
system is a function $L: T\mbox {\rm Diff}(\mathcal{D}) \times V^\ast
\rightarrow \mathbb{R}$ which is right invariant relative to the
tangent lift of right translation of $\mbox {\rm Diff}(\mathcal{D})$
on itself and pull back on the tensor field densities.

Thus, the Lagrangian $L$ induces
a {\it reduced La\-gran\-gian} $l: \mathfrak{X}(\mathcal{D}) \times
V^\ast \rightarrow \mathbb{R}$ defined by
\[
l(\mathbf{u}, a) = L(\mathbf{u}\circ \eta, \eta^\ast a),\] where $
\mathbf{u} \in \mathfrak{X}({\mathcal{D}})$ and $  a \in V^\ast
\subset {\mathfrak{T}}({\mathcal{D}})\otimes {\rm
Den}({\mathcal{D}})$, and where $\eta^\ast a$ denotes the pull
back of $a$ by the diffeomorphism $\eta$ and $\mathbf{u}$ is the
Eulerian velocity. The evolution of $a$ is given by solving the
equation
\[
\dot a = -{\pounds}_\mathbf{u}\, a.
\]
The solution of this equation, given the
initial condition $a_0$, is $a(t) = \varphi_{t\ast} a_0$,
where the lower star denotes the push forward operation and
$\varphi_t$ is the flow of $\mathbf{u}$.

\paragraph{Advected Eulerian Quantities.} These are defined in
continuum mechanics to be those variables which are Lie transported
by the flow of the Eulerian velocity field. Using this standard
terminology, the above equation states that the tensor field density
$a$ (which may include mass density and other Eulerian quantities)
is advected.

As we have discussed, in many examples, $V^\ast \subset
{\mathfrak{T}}({\mathcal{D}})\otimes {\rm Den}({\mathcal{D}})$.  On a
general manifold, tensors of a given type have natural duals. For
example, symmetric covariant tensors are dual to symmetric
contravariant tensor densities, the pairing being given by the
integration of the natural contraction of these tensors. Likewise,
$k$--forms are naturally dual to
$(n-k)$--forms, the pairing being given by taking the integral of
their wedge product.

The operation $\diamond$ between elements of $V$ and
$V^\ast$  producing an element of  $\mathfrak{X}({\mathcal{D}
})^\ast$ introduced in equation (\ref{def-diamond}) becomes
\begin{equation}\label{continuumdiamond}
\langle v \diamond a, \mathbf{w}\rangle
= \int_{\mathcal{D}} a
\cdot \pounds_\mathbf{w}\,v
= -\int_{\mathcal{D}} v
\cdot \pounds_\mathbf{w}\,a
\;,
\end{equation} where $v\cdot \pounds_\mathbf{w}\,a$ denotes the
contraction, as described above, of elements of $V$ and elements of
$V^\ast$. (These operations do {\it not} depend on a Riemannian
structure.)

For a path $\eta_t \in {\rm Diff}(\mathcal{D})$ let $\mathbf{u}(x,
t)$ be its Eulerian velocity and consider, as in the hypotheses of
Theorem \ref{rarl} the curve $a(t)$ with initial condition $a_0$
given by the equation
\begin{equation}
\dot a + \pounds_\mathbf{u} a = 0.
\label{continuityequation}
\end{equation}
Let $L_{a_0}(\mathbf{U} ) := L(\mathbf{U} , a_0)$.
We can now state Theorem \ref{rarl} in this particular, but
very useful, setting.

\begin{thm}[Euler--Poincar\'e reduction for continua.]
\label{EPforcontinua}
For a path $\eta_t$ in  ${\rm Diff}(\mathcal{D})$ with
Lagrangian velocity $\mathbf{U} $ and Eulerian velocity $\mathbf{u}$,
the following are equivalent:

\begin{enumerate}
\item [{\bf i}] Hamilton's variational principle
\begin{equation} \label{continuumVP}
\delta \int_{t_1}^{t_2} L\left(X, \mathbf{U}_t (X),
a_0(X)\right)\,dt=0
\end{equation}
holds, for variations $\delta\eta_t$ vanishing at the endpoints.
\item [{\bf ii}] $\eta_t$ satisfies the Euler--Lagrange
equations for $L_{a_0}$ on ${\rm Diff}(\mathcal{D})$.
\item [{\bf iii}] The constrained variational principle in
Eulerian coordinates
\begin{equation}\label{continuumconstrainedVP}
   \delta \int_{t_1}^{t_2} l(\mathbf{u},a)\ dt=0
\end{equation}
holds on $\mathfrak{X}(\mathcal{D}) \times V^\ast$, using
variations of the form
\begin{equation}\label{continuumvariations}
   \delta \mathbf{u} = \frac{\partial \mathbf{w}}{\partial t}
                      -\, {\rm ad}_\mathbf{u}\mathbf{w}
                     = \frac{\partial \mathbf{w}}{\partial t}
                   + [\mathbf{u},\mathbf{w}], \quad
   \delta a = - \pounds_\mathbf{w}\,a,
\end{equation}
where $\mathbf{w}_t = \delta\eta_t \circ \eta_t^{-1}$ vanishes at
the endpoints.
\item [{\bf iv}] The {\bfi Euler--Poincar\'{e} equations for continua}
\begin{equation}\label{continuumEP}
   \frac{\partial }{\partial t}\frac{\delta l}{\delta \mathbf{u}}
   = -\, {\rm ad}^{\ast}_\mathbf{u}\frac{\delta l}{\delta \mathbf{u}}
   +\frac{\delta l}{\delta a}\diamond a
   =-\pounds_\mathbf{u} \frac{\delta l}{\delta \mathbf{u}}
    +\frac{\delta l}{\delta a}\diamond a\,,
\end{equation}
hold, where the $\diamond$ operation given by
(\ref{continuumcoadjoint}) needs to be determined on a
case by case basis, depending on the nature of the tensor $a$.
(Remember that $\delta l/\delta \mathbf{u}$ is a one--form density.)
\end{enumerate}
\end{thm}

\paragraph{Remarks.}
\begin{enumerate}
\item The following string of equalities
shows {\it directly} that {\bf iii} is equivalent to {\bf iv}:
\begin{eqnarray}\label{continuumEPderivation}
0
&=&\delta \int_{t_1}^{t_2} l(\mathbf{u}, a) dt
=\int_{t_1}^{t_2}\left(\frac{\delta l}{\delta \mathbf{u}}\cdot
\delta\mathbf{u} +\frac{\delta l}{\delta a}\cdot \delta a\right)dt
\nonumber \\
&=&\int_{t_1}^{t_2} \left[\frac{\delta l}{\delta \mathbf{u}}
\cdot \left(\frac{\partial\mathbf{w}}
{\partial t}-{\rm ad}_\mathbf{u}\,\mathbf{w}\right) -\frac{\delta
l}{\delta a}\cdot \pounds_\mathbf{w}\, a \right]dt
\nonumber \\
&=&\int_{t_1}^{t_2} \mathbf{w}\cdot
\left[-\,\frac{\partial}{\partial t}
\frac{\delta l}{\delta\mathbf{u}} -{\rm ad}^*_\mathbf{u}\frac
{\delta l}{\delta\mathbf{u}} +\frac{\delta l}{\delta a} \diamond
a\right]dt\,.
\end{eqnarray}

\item Similarly, one can deduce the form (\ref{continuumvariations})
of the variations in the constrained variational principle
(\ref{continuumconstrainedVP}) by a direct calculation, as follows.
One writes the basic relation between the spatial and material
velocities, namely
$ \mathbf{u} (x,t) = \dot{\eta} (\eta_t ^{-1} (x),t) $. One then
takes the variation of this equation with respect to $\eta$ and uses
the definition $ \mathbf{w} (x,t) = \delta \eta ((\eta_t ^{-1}
(x),t) $ together with a calculation of its time derivative. Of
course, one can also do this calculation using the inverse map
$\eta _t ^{-1} $ instead of $\eta$ as the basic variable, see Holm,
Marsden, and Ratiu [1986], Holm [1996].

\item The preceding sort of calculation for $\delta \mathbf{u}$ in
fluid mechanics and the interpretation of this restriction on the
form of the variations as the so-called Lin constraints is due to
Bretherton [1970]. The variational form (\ref{continuumvariations})
for the `basic' Euler--Poincar\'e equations (i.e., without the
advected parameters $a$) is due to Marsden and Scheurle [1993a] and
Bloch, Krishnaprasad, Marsden and Ratiu [1996]. For the full
Euler--Poincar\'e case, this form is due to Holm, Marsden and Ratiu
[1998a] and for the general Lagrangian reduction case, to Cendra, Holm,
Marsden and Ratiu [1997] and Cendra, Marsden and Ratiu [1997]. These
ideas were used for Maxwell-Vlasov plasmas by Cendra, Holm, Hoyle and
Marsden [1997].

\item The coordinate expressions for
$({\delta l/ \delta a})\diamond a$ required to complete the
equations of motion for GFD models are given in the next section for
$a_0(X)$ in three dimensions.

\item As with the general theory, variations of the action in the
advected tensor quantities contribute to the equations of motion
which follow from Hamilton's principle. At the level of the action
$l$ for the Euler-Poincar\'{e} equations, the Legendre transform in
the variable $\mathbf{u}$ alone is often nonsingular, and when it
is, it produces the Hamiltonian formulation of Eulerian fluid
motions with a Lie-Poisson bracket defined on the dual of the
semidirect product algebra of vector fields acting amongst
themselves by Lie bracket and on tensor fields and differential
forms by the Lie derivative. This is a special instance of the more
general facts for Lie algebras that was discussed earlier.
\end{enumerate}

\paragraph{The Inverse Map and the Tensor Fields $a$ for
Fluids.} In the case of fluids in the Lagrangian picture, the flow
of the fluid is a diffeomorphism which takes a fluid parcel along a
path from its initial position $X$, in a ``reference configuration"
to its current position $x$ in the ``container", i.e., in the
Eulerian domain of flow. As we have described, this {\bfi forward map}
is denoted by $\eta : X \mapsto x$. The inverse map $\eta^{-1}: x
\mapsto X$ provides the map assigning the Lagrangian labels to a
given spatial point. Interpreted as passive scalars, these
Lagrangian labels are simply advected with the fluid flow, $\dot
X=0$. In the Lagrangian picture, a tensor density in the reference
configuration $a_0(X)$ (satisfying $\dot a_0(X)=0$) consists of {\it
invariant} tensor functions of the initial reference  positions and
their differentials. These tensor functions are parameters of the
initial fluid reference configuration (e.g., the  initial density
distribution, which is an invariant $n$-form).

When viewed in the Eulerian picture as
\[ a_t (x):= (\eta _{t \ast}  a_0)(x)
= (\eta^{-1 \ast}_t  a_0)(x),\]
{\rm i.e.\/},
\[ a_0 (X):= (\eta ^\ast _{t}  a_t)(X)
= (\eta^{-1}_{t \ast} a_0)(X),\]
the time invariant tensor density $a_0(X)$ in
the reference configuration acquires advective dynamics in the
Eulerian picture, namely
\[ \dot a_0(X) = \left(\frac{\partial}{\partial t} +
\pounds_\mathbf{u}\right)\,a(x,t)=0,\]  where $\pounds_\mathbf{u}$
denotes Lie derivative with respect to the Eulerian velocity field
$\mathbf{u}(x,t)$. This relation results directly from the
well known Lie derivative formula for tensor fields. (See, for
example, Abraham, Marsden and Ratiu [1988].)

Mapping the time invariant quantity $ a _0 $ (a tensor density
function of $X$) to the time varying quantity $ a _t $ (a tensor
density function of $x$) as explained above is a special case of
the way we advect quantities in $ V ^\ast $ in the general
theory. Specifically, we can view this advection of $ a _t $ as
being the fluid analogue of the advection of the unit vector along
the direction of gravity (a spatially fixed quantity) by means of the
body rotation vector in the heavy top example.

Consistent with the fact that the heavy top is a {\it left invariant}
system while continuum theories are {\it right invariant}, the
advected tensor density $a_t$ is a spatial quantity, while the
advected direction of gravity is a body quantity. If we were to take
the inverse map $\eta^{-1}$ as the basic group variable, rather than
the map $\eta $, then continuum theories would also become left
invariant.

\paragraph{The Continuity Equation for the Mass Density.} We will
need to make an additional assumption on our continuum theory.
Namely, we assume that amongst the tensor densities being advected,
there is a special one, namely the mass density. This of course is a
tensor density that occurs in all continuum theories. We denote this
density by $\rho d^n x$ and it is advected according to the standard
principles discussed above. Thus, $\rho$ satisfies the usual
continuity equation:
\[
    \frac{\partial }{\partial t} \rho
    + {\rm div} (\rho \mathbf{u} ) = 0.
\]
In the Lagrangian picture we have $\rho d^n x=\rho_0(X) d^n X$, where
$\rho_0(X)$ is the mass density in the reference configuration. It
will also be convenient in the continuum examples below to define
Lagrangian {\it mass} coordinates $\ell(X)$ satisfying $\rho d^n
x=d^n \ell$ with $\dot \ell = 0$. (When using Lagrangian mass
coordinates, we shall denote the density $\rho$ as $D$.)

\paragraph{The Kelvin--Noether Theorem.} Let
\[
\mathcal C :=  \left. \left\{ \gamma : S^1 \rightarrow {\mathcal{D}}
\; \right| \; \gamma {\rm~continuous} \right\}
\]
 be the space of continuous loops in $\mathcal{D}$ and
let the group  $\mbox {\rm Diff}(\mathcal{D})$ act on
$\mathcal C$ on the left by $(\eta, \gamma)\in \mbox {\rm
Diff}(\mathcal{D}) \times {\mathcal C} \mapsto \eta\gamma \in
{\mathcal C}$, where $\eta\gamma = \eta\circ\gamma$.

Next we shall define the {\bfi circulation map\/}
$\mathcal K: {\mathcal C} \times V ^\ast
\rightarrow \mathfrak{X}(\mathcal{D})^{\ast\ast}$.
Given a one form density $\alpha \in \mathfrak{X} ^\ast $ we can
form a one form (no longer a density) by dividing it by the mass
density $\rho$; we denote the result just by $\alpha  / \rho$.
We let $ \mathcal{K} $ then be defined by
\begin{equation}
 \left\langle \mathcal{K} (\gamma , a) , \alpha
\right\rangle  = \oint _\gamma \frac{ \alpha }{ \rho } \, .
\label{circ.map}
\end{equation}
The expression in this definition is called the {\bfi circulation\/}
of the one--form $\alpha/\rho$ around the loop $\gamma$.

This map is equivariant in the sense that
\[ \left\langle \mathcal{K} ( \eta \circ \gamma , \eta _\ast a) ,
\eta _\ast \alpha \right\rangle =
\left\langle \mathcal{K} (\gamma, a), \alpha \right\rangle
\]
for any $ \eta \in {\rm Diff} ( {\mathcal{D}} ) $. This is proved
using the definitions and the change of variables formula.

Given the Lagrangian $l:\mathfrak{X}(\mathcal{D}) \times V^\ast
\rightarrow {\mathbb{R}}$, the Kelvin--Noether quantity is given
by (\ref{KelvinNoether}) which in this case becomes
\begin{equation} \label{KN-quantity}
I({\gamma_t}, \mathbf{u}, a) = \oint_{\gamma_t}
\frac{1}{\rho}\frac{\delta l}{\delta \mathbf{u}}\;.
\end{equation}
With these definitions of $\mathcal K$ and $I$, the statement of
Theorem \ref{KelvinNoetherthm} becomes the classical Kelvin
circulation theorem.

\begin{thm}[Kelvin Circulation Theorem.] \label{KelThmforcontinua}
Assume that
$\mathbf{u}(x, t)$ satisfies the Euler--Poincar\'e equations for
continua (\ref{continuumEP}):
\[
\frac{\partial }{\partial t}\left(\frac{\delta
l}{\delta \mathbf{u}}\right)
   = -\pounds_\mathbf{u}
   \left(\frac{\delta l}{\delta \mathbf{u}}\right)
    +\frac{\delta l}{\delta a}\diamond a
\]
and $a$ satisfies
\[\frac{\partial a}{\partial t} +
\pounds_\mathbf{u} a = 0.
\]
Let $\eta_t$ be the flow of the Eulerian
velocity field $\mathbf{u}$, that is, $\mathbf{u}_t  =
(d\eta_t/dt)\circ \eta_t^{-1}$. Define
$\gamma_t  := \eta_t\circ \gamma_0$ and $I(t)
:= I(\gamma_t , \mathbf{u}_t , a_t)$. Then
\begin{equation} \label{KN-theorem}
\frac{d}{dt}I(t) = \oint_{\gamma_t }
\frac{1}{\rho}\frac{\delta l}{\delta a}\diamond a\;.
\end{equation}
\end{thm}
In this statement, we use a subscript $t$ to emphasise that the
operations are done at a particular $t$ and to avoid having to write
the other arguments, as in $ a _t (x) = a(x,t)$; we omit the
arguments from the notation when convenient. Due to the importance
of this theorem we shall give here a separate proof tailored for the
case of continuum mechanical systems.
\medskip

\noindent{\bf Proof.\,} First we change variables in the expression
for $I(t)$:
\[
I(t) = \oint_{\gamma_t}\frac{1}{\rho_t}
\frac{\delta l}{\delta \mathbf{u}}
=\oint_{\gamma_0} \eta_t^\ast\left[\frac{1}{\rho_t}\frac{\delta l}
{\delta \mathbf{u}}\right] = \oint_{\gamma_0}
\frac{ 1 }{ \rho _0 } \eta_t^\ast\left[\frac{\delta
l} {\delta \mathbf{u}}\right].
\]
Next, we use the Lie derivative formula, namely
\[
\frac{d}{dt}\left(\eta_t^*\alpha_t\right) =
\eta_t^*\left(\frac{\partial}{\partial t}\alpha_t
+ \pounds_\mathbf{u} \alpha_t \right)\;,
\]
applied to a one--form density $\alpha_t$.
This formula gives
\begin{eqnarray*}
      \frac{d}{dt} I(t)
  & = & \frac{d}{dt}  \oint_{\gamma_0}
\frac{ 1 }{ \rho _0 } \eta_t^\ast\left[\frac{\delta
l} {\delta \mathbf{u}}\right] \\
  & = & \oint_{\gamma_0} \frac{1}{\rho _0} \frac{d}{dt}
\left( \eta_t^\ast\left[
\frac{\delta l} {\delta \mathbf{u}}\right]\right)  \\
  & = & \oint_{\gamma_0} \frac{1}{\rho _0} \eta_t^*\left[
\frac{\partial}{\partial t}
\left(\frac{\delta l} {\delta \mathbf{u}}\right) +
\pounds_\mathbf{u}
\left(\frac{\delta l} {\delta \mathbf{u}} \right)\right].
\end{eqnarray*}
By the Euler--Poincar\'e equations (\ref{continuumEP}), this becomes
\[
      \frac{d}{dt} I(t)
  =   \oint_{\gamma_0} \frac{1}{\rho _0} \eta_t^*\left[
\frac{\delta  l}{\delta  a}
\diamond a \right] = \oint_{\gamma_t} \frac{1}{\rho _t} \left[
\frac{\delta  l}{\delta  a}
\diamond a \right],
\]
again by the change of variables formula.\quad $\blacksquare$

\begin{cor}
Since the last expression holds for every loop $\gamma_t$, we may
write it as
\begin{equation}
\left(\frac{\partial}{\partial t} + \pounds_\mathbf{u} \right)
\frac{1}{\rho} \frac{\delta l}{\delta \mathbf{u}}
= \frac{1}{\rho} \frac{\delta  l}{\delta  a} \diamond a\,.
\label{KThfm}
\end{equation}
\end{cor}

\section{Applications of the Euler--Poincar\'{e} Theorem in GFD}
\label{sec-GFD-applic}

\paragraph{Variational Formulae in Three Dimensions.} We compute
explicit formulae for the variations $\delta a$ in the cases that
the set of tensors $a$ is drawn from a set of scalar fields and
densities on ${\mathbb{R}}^3$. We shall denote this symbolically by
writing
\begin{equation}
a\in\{b,D\,d^3x\}\,.
\label{Eul-ad-qts}
\end{equation}
We have seen that invariance of the
set $a$ in the Lagrangian picture under the dynamics of
$\mathbf{u}$ implies in the Eulerian picture that
\[
     \left( \frac{\partial}{\partial t}
         + \pounds_\mathbf{u} \right) \,a=0 \, ,
\]
where $\pounds_\mathbf{u}$ denotes Lie derivative with respect to the
velocity vector field $\mathbf{u}$. Hence, for a fluid dynamical
action $\mathfrak{S}=\int\,dt\ l(\mathbf{u};b,D)$, the advected
variables
$\{b, D\}$ satisfy the following Lie-derivative relations,
\begin{eqnarray}
\left(\frac{\partial}{\partial t}+ \pounds_\mathbf{u}\right) b=0,
&{\rm or}&
\frac{\partial b}{\partial t}
= -\ \mathbf{u}\cdot\nabla\,b\,,
\label{eqn-b} \\
\left(\frac{\partial}{\partial t}+ \pounds_\mathbf{u}\right)D\,d^3x=0,
&{\rm or}&
\frac{\partial D}{\partial t}
= -\ \nabla\cdot(D\mathbf{u})\,.
\label{eqn-D}
\end{eqnarray}
In fluid dynamical applications, the advected Eulerian variables $b$
and $D$ represent the buoyancy $b$ (or specific entropy, for the
compressible case) and volume element (or mass density) $D$,
respectively. According to Theorem \ref{EPforcontinua}, equation
(\ref{continuumconstrainedVP}), the variations of the tensor
functions $a$ at fixed $\mathbf{x}$ and $t$ are also given by Lie
derivatives, namely
$\delta a = - \pounds_\mathbf{w}\,a$, or
\begin{eqnarray}
\delta b
&=& -\pounds_\mathbf{w}\ b = -\mathbf{w}\cdot\nabla\,b\,,
\nonumber \\
\delta D\ d^3x&=&-\pounds_\mathbf{w}\,(D\,d^3x)
= -\nabla\cdot(D\mathbf{w})\ d^3x
\,.
\end{eqnarray}
Hence, Hamilton's principle with this dependence yields
\begin{eqnarray}
0 &=&\delta \int dt\ l(\mathbf{u}; b,D)
\nonumber \\
&=&\int dt\ \bigg[\frac{\delta l}{\delta
\mathbf{u}}\cdot \delta\mathbf{u}
+\frac{\delta l}{\delta b}\ \delta b
+\frac{\delta l}{\delta D}\ \delta D
\bigg]
\nonumber \\
&=&\int dt\ \bigg[\frac{\delta l}{\delta \mathbf{u}}
\cdot \Big(\frac{\partial \mathbf{w}}{\partial t}
-{\rm ad}_\mathbf{u}\,\mathbf{w}\Big)
-\frac{\delta l}{\delta b}\ \mathbf{w}\cdot\nabla\,b
-\frac{\delta l}{\delta D}\ \Big(\nabla
\cdot(D\mathbf{w})\Big)\bigg]
\nonumber \\
&=&\int dt\ \mathbf{w}\cdot
\bigg[-\frac{\partial }{\partial t}
\frac{\delta l}{\delta \mathbf{u}}
-{\rm ad}^*_\mathbf{u}\ \frac{\delta l}{\delta \mathbf{u}}
-\frac{\delta l}{\delta b}\ \nabla\,b
+D\ \nabla\frac{\delta l}{\delta D}\bigg]
\nonumber \\
&=&-\int dt\ \mathbf{w}\cdot
\bigg[\Big(\frac{\partial }{\partial t}
 +\pounds_\mathbf{u} \Big)\frac{\delta l}{\delta \mathbf{u}}
+\frac{\delta l}{\delta b}\ \nabla\,b
-D\ \nabla\frac{\delta l}{\delta D}\bigg]\,,
\label{eq-EP-Eul}
\end{eqnarray}
where we have consistently dropped boundary terms arising from
integrations by parts, by invoking natural boundary conditions.
Specifically, we impose $\hat\mathbf{n}\cdot\mathbf{w}=0$ on the
boundary, where $\hat\mathbf{n}$ is the boundary's outward unit normal
vector.

The Euler--Poincar\'e equations for continua (\ref{continuumEP}) may
now be summarized for advected Eulerian variables $a$ in the set
(\ref{Eul-ad-qts}). We adopt the notational convention of the
circulation map $\mathcal K$ in equation (\ref{circ.map}) that a one
form density can be made into a one form (no longer a density) by
dividing it by the mass density $D$ and use the Lie-derivative
relation for the continuity equation
$({\partial}/{\partial t}+\pounds_\mathbf{u})Dd^3x = 0$. Then, the
Euclidean components of the Euler--Poincar\'e equations for continua
in equation (\ref{eq-EP-Eul}) are expressed in Kelvin theorem form
(\ref{KThfm}) with a slight abuse of notation as
\begin{equation}
\Big(\frac{\partial }{\partial t} + \pounds_\mathbf{u}\Big)
\Big(\frac{1}{D}\frac{\delta l}{\delta
\mathbf{u}}\cdot d\mathbf{x}\Big)
\,+\,\frac{1}{D}\frac{\delta l}{\delta b}\nabla b \cdot d\mathbf{x}
\,-\,\nabla\Big(\frac{\delta l}{\delta D}\Big)\cdot d\mathbf{x}
 = 0\,,
\label{EP-Kthm}
\end{equation}
in which the variational derivatives of the Lagrangian $l$ are to be
computed according to the usual physical conventions, i.e., as
Fr\'echet derivatives. Formula (\ref{EP-Kthm}) is the
Kelvin--Noether form of the equation of motion for ideal continua.
Hence, we have the explicit Kelvin theorem expression, cf. equations
(\ref{KN-quantity}) and (\ref{KN-theorem}),
\begin{equation} \label{KN-theorem-bD}
\frac{d}{dt}\oint_{\gamma_t(\mathbf{u})} \frac{1}{D}\frac{\delta
l}{\delta \mathbf{u}}\cdot d\mathbf{x}
= -\oint_{\gamma_t(\mathbf{u})}
\frac{1}{D}\frac{\delta l}{\delta b}\nabla b \cdot d\mathbf{x}\;,
\end{equation}
where the curve $\gamma_t(\mathbf{u})$ moves with the fluid velocity
$\mathbf{u}$. Then, by Stokes' theorem, the Euler equations generate
circulation of $(\frac{1}{D}\frac{\delta{l}}{\delta\mathbf{u}})$
whenever~$\nabla b$ and $\nabla(\frac{1}{D}\frac{\delta l}{\delta
b})$ are not collinear. The corresponding conservation of potential
vorticity $q$ on fluid parcels is given by
\begin{equation} \label{pv-cons-EP}
\frac{\partial{q}}{\partial{t}}+\mathbf{u}\cdot\nabla{q} =
0\,,
\quad \hbox{where}\quad
{q}=\frac{1}{D}\nabla{b}\cdot{\rm curl}
\left(\frac{1}{D}\frac{\delta l}{\delta \mathbf{u}}\right).
\end{equation}

\paragraph{Remark on Lagrangian Mass Coordinates.} An alternative way
to treat Hamilton's principle for an action $\mathfrak{S}=\int\,dt\
l(\mathbf{u};b,D)$ is to perform variations at fixed $\mathbf{x}$
and $t$ of the {\it inverse} maps $\mathbf{x}\rightarrow
{\mbox{\boldmath$\ell$}}$, described by the Lagrangian mass
coordinate functions $\ell^A(\mathbf{x},t)$,
$A=1,2,\dots,n$, which determine $\mathbf{u}$, $b$ and $D$ by the
formulae (in which one sums on repeated indices)
\begin{equation}
\frac{\partial \ell^A}{\partial t} = -u^iD^A_i\,,
\quad b=b(\ell^A)\,,
\quad D^A_i=\frac{\partial \ell^A}{\partial x^i} \,,
\quad D=\mathrm{det}(D^A_i)\,.
\label{lag-def}
\end{equation}
As discussed above, the relation of mass coordinates $\ell$ to the
usual Lagrangian coordinates $X$ is given by a simple change of
variables in the fluid reference configuration to make
$\rho_0(X)d^nX=d^n\ell$. Variation of an action of the form
$\mathfrak{S}=\int\,dt\ l(\mathbf{u},b,D)$ with respect to
$\ell^A$ with $p$ imposing volume preservation then yields (Holm,
Marsden, and Ratiu [1986], Holm [1996]),
\begin{eqnarray}
{\delta}\mathfrak{S} &=& \int dt \int d^nx\
\Big\{ D (D^{-1})^i_A {\delta}\ell^A
\Big[\frac{d}{dt} \frac{1}{D} \frac{{\delta} l}{{\delta} u^i}
\,+\, \frac{1}{D} \frac{{\delta} l}{{\delta} u^j} u^j_{,i}
\nonumber \\
&&\hspace{1in}
+\, \frac{1}{D} \frac{{\delta} l}{{\delta} b} b_{,i}
- \Big(\frac{{\delta} l}{{\delta} D}\Big)_{,i}\,\Big]
-\, {\delta} p(D-1)\Big\}\,,
\label{hpg}
\end{eqnarray}
where $d/dt=\partial/\partial t+\mathbf{u}\cdot\nabla$ is the
material derivative and we again invoke natural boundary conditions
($\hat\mathbf{n}\cdot\mathbf{u}=0$ on the boundary) when integrating
by parts. Hence, the vanishing of the coefficient of
${\delta}\ell^A$ in the variational formula (\ref{hpg}) recovers the
Euler--Poincar\'e equations for continua (\ref{continuumEP})
expressed in Kelvin theorem form (\ref{KThfm}) or
(\ref{KN-theorem-bD}), by stationarity of the action
$\mathfrak{S}=\int\,dt\ l(\mathbf{u};b,D)$ with respect to variations
of the Lagrangian mass coordinates
$\ell^A(\mathbf{x},t)$. In vector notation, these equations are
\begin{equation}
\frac{d}{dt} \frac{1}{D} \frac{{\delta} l}{{\delta} \mathbf{u}}
\,+\, \frac{1}{D} \frac{{\delta} l}{{\delta} u^j} \nabla u^j
\,+\, \frac{1}{D} \frac{{\delta} l}{{\delta} b} \nabla b
-  \nabla\frac{{\delta} l}{{\delta} D}=0,
\label{EP-comp1}
\end{equation}
or, in three dimensions,
\begin{equation}
\frac{\partial}{\partial t}
\Big(\frac{1}{D} \frac{{\delta} l}{{\delta} \mathbf{u}}\Big)
\,-\, \mathbf{u}\times {\rm curl}
\Big(\frac{1}{D} \frac{{\delta} l}{{\delta} \mathbf{u}}\Big)
\,+\, \nabla\Big(\mathbf{u}\cdot\frac{1}{D}
\frac{{\delta} l}{{\delta} \mathbf{u}}
\,-\, \frac{{\delta} l}{{\delta} D}\Big)
\,+\, \frac{1}{D} \frac{{\delta} l}{{\delta} b} \nabla b
=0\,.
\label{EP-comp2}
\end{equation}
In writing the last equation, we have used the fundamental vector
identity of fluid dynamics,
\begin{equation}
(\mathbf{b}\cdot\nabla)\mathbf{a} + a_j\nabla
b^j =-\ \mathbf{b}\times(\nabla\times \mathbf{a})
+ \nabla(\mathbf{b}\cdot\mathbf{a})\,,
\label{fvid}
\end{equation}
for any three dimensional vectors $\mathbf{a}$ and $\mathbf{b}$
with, in this case, $\mathbf{a} = (\frac{1}{D}
\frac{\delta{l}}{\delta\mathbf{u}})$ and $\mathbf{b}=\mathbf{u}$.
Taking the curl of equation (\ref{EP-comp2}) and using advection of the
buoyancy $b$ yields conservation of potential vorticity on fluid
parcels as given in equation (\ref{pv-cons-EP}). For incompressible
flows $D=1$ in equation (\ref{pv-cons-EP}). The Euclidean component
formulae (\ref{EP-comp1}), (\ref{EP-comp2}) and (\ref{pv-cons-EP})
are especially convenient for direct calculations of motion equations
in geophysical fluid dynamics, to which we turn our attention next.

\subsection{Euler's Equations for a Rotating Stratified Ideal
Incompressible Fluid}
\label{sec-Euler}

\paragraph{The Lagrangian.} In the Eulerian velocity representation, we
consider Hamilton's principle for fluid motion in an three dimensional
domain with action functional $\mathfrak{S}=\int\,dt\, l$ and
Lagrangian
$l(\mathbf{u},b,D)$ given by
\begin{equation}
l = \int d^{\,3}x\
\rho_0 D (1+b)
\Big(\frac{1}{2} |\mathbf{u}|^2
+ \mathbf{u}\cdot\mathbf{R}(\mathbf{x}) - gz\Big)
 - p(D-1)\,,
\label{lag-v}
\end{equation}
where $\rho_{tot}=\rho_0 D (1+b)$ is the total mass density, $\rho_0$
is a dimensional constant and $\mathbf{R}$ is a given function of
$\mathbf{x}$. This Lagrangian produces the following variations at
fixed $\mathbf{x}$ and $t$
\begin{eqnarray}
&&\frac{1}{D}\frac{{\delta} l}{{\delta} \mathbf{u}}
= \rho_0(1+b)(\mathbf{u}+ \mathbf{R})\,,
\quad
\frac{{\delta} l}{{\delta} b}
= \rho_0 D\Big(\frac{1}{2}|\mathbf{u}|^2
+ \mathbf{u}\cdot\mathbf{R} - gz\Big)\,,
\nonumber \\
&&\frac{{\delta} l}{{\delta} D}
= \rho_0(1+b)\Big(\frac{1}{2}|\mathbf{u}|^2
+ \mathbf{u}\cdot\mathbf{R} -
gz\Big) - p\,,
\quad
\frac{{\delta} l}{{\delta} p} = -\,(D-1)\,.
\hspace{.25in}
\label{vds-1}
\end{eqnarray}
Hence, from the Euclidean component formula (\ref{EP-comp1}) for
Hamilton principles of this type and the fundamental vector identity
(\ref{fvid}), we find the motion equation for an Euler fluid in
three dimensions,
\begin{equation}
\frac{d\mathbf{u}}{dt} - \mathbf{u} \times {\rm curl}\, \mathbf{R}
+g\hat{\bf z}
+ \frac{1}{\rho_0(1+b)}\nabla p = 0\,,
\label{Eul-mot}
\end{equation}
where ${\rm curl}\,\mathbf{R}=2\boldsymbol{\Omega}(\mathbf{x})$ is the
Coriolis parameter (i.e., twice the local angular rotation
frequency). In writing this equation, we have used advection of
buoyancy,
$$
\frac{\partial{b}}{\partial{t}}+\mathbf{u}\cdot\nabla{b} = 0,
$$
from equation (\ref{eqn-b}).

\paragraph{The Kelvin--Noether Theorem.} From equation
(\ref{KN-theorem-bD}), the Kelvin--Noether circulation theorem
corresponding to the motion equation (\ref{Eul-mot}) for an ideal
incompressible stratified fluid in three dimensions is,
\begin{equation}
\frac{d}{dt}\oint_{\gamma_t(\mathbf{u})}(\mathbf{u}
+\mathbf{R})\cdot d\mathbf{x}
= -\oint_{\gamma_t(\mathbf{u})}
\frac{1}{\rho_0(1+b)}\nabla p\cdot d\mathbf{x}\;,
\label{KN-theorem-Eul}
\end{equation}
where the curve $\gamma_t(\mathbf{u})$ moves with the fluid velocity
$\mathbf{u}$. By Stokes' theorem, the Euler equations generate
circulation of $(\mathbf{u}+\mathbf{R})$ around $\gamma_t(\mathbf{u})$
whenever the gradients of bouyancy and pressure are not collinear.
Using advection of buoyancy $b$, one finds conservation of potential
vorticity $q_{\rm Eul}$ on fluid parcels, cf. equation
(\ref{pv-cons-EP}),
\begin{equation} \label{pv-cons-Eul}
\frac{\partial{q}_{\rm Eul}}{\partial{t}}+\mathbf{u}\cdot\nabla{q}_{\rm Eul} =
0\,,
\quad \hbox{where}\quad
{q}_{\rm Eul}=\nabla{b}\cdot{\rm curl}\,(\mathbf{u}+\mathbf{R})\,.
\end{equation}

The constraint $D=1$ (volume preservation) is imposed by varying $p$
in Hamilton's principle, according to equation (\ref{vds-1}).
Incompressibility then follows from substituting $D=1$ into the
Lie-derivative relation (\ref{eqn-D}) for $D$, which gives
$\nabla\cdot\mathbf{u}=0$.  Upon taking the divergence of the motion
equation (\ref{Eul-mot}) and requiring incompressibility to be
preserved in time, one finds an elliptic equation for the pressure
$p$ with a Neumann boundary condition obtained from the normal
component of the motion equation (\ref{Eul-mot}) evaluated on the
boundary.

\subsection{Euler-Boussinesq Equations} \label{sec-EB}

\paragraph{The Lagrangian.} The Lagrangian (\ref{lag-v}) for the Euler
fluid motion nondimensionalizes as follows, in terms of units of $L$
for horizontal distance, $B_0$ for vertical depth, ${\cal U}_0$ for
horizontal velocity,  $B_0{\cal U}_0/L$ for vertical velocity, $f_0$
for Coriolis parameter, $\rho_0$ for density and
$\rho_0f_0L{\cal U}_0$ for pressure:
\begin{equation}
l = \int d^{\,3}x\ D (1+b)
\Big(\frac{\epsilon}{2}\mathbf{u}_3\cdot\mathbf{v}_3
+ \mathbf{u}\cdot\mathbf{R}(\mathbf{x}) -
\frac{z}{\epsilon{\cal F}}\Big)
 - p(D-1)\,.
\label{lag-Eul}
\end{equation}
Here we take $\mathbf{R}(\mathbf{x})$ to be horizontal and
independent of the vertical coordinate, so ${\rm
curl}\,\mathbf{R}=f(\mathbf{x})\hat{\mbox{\boldmath{$z$}}}$. In this
nondimensional notation, three-dimensional vectors and gradient
operators have subscript 3, while horizontal vectors and gradient
operators are left unadorned. Thus, we denote, in three dimensional
Euclidean space,
\begin{eqnarray}
&&\hspace{0.4in}\mathbf{x}_3 =(x,y,z), \quad \mathbf{x} =(x,y,0),
\nonumber \\
&&\hspace{0.3in}\mathbf{u}_3 = (u,v,w), \quad \mathbf{u} = (u,v,0),
\nonumber \\
&&\nabla_3 = \left(
\frac{\partial}{\partial{x}},\frac{\partial}{\partial{y}},
\frac{\partial}{\partial{z}}
\right),
\quad \nabla = \left(
\frac{\partial}{\partial{x}},\frac{\partial}{\partial{y}},0\right),
\nonumber \\
&&\quad
\frac{ d}{ dt} = \frac{\partial}{\partial{t}}
+ \mathbf{u}_3\cdot\nabla_3
= \frac{\partial}{\partial{t}} + \mathbf{u}\cdot\nabla
+ w \frac{\partial}{\partial{z}}\ ,
\label{3-notation}
\end{eqnarray}
and $\hat{\mbox{\boldmath{$z$}}}$ is the unit vector in the vertical
$z$ direction.  For notational convenience, also we define {\it two}
nondimensional velocities
\begin{equation} \label{uv-def}
\mathbf{u}_3 = (u,v,w),
\quad \hbox{and} \quad
\mathbf{v}_3 = (u,v,\sigma^2 w)\,,
\end{equation}
as well as three nondimensional parameters,
\begin{equation}
\epsilon = \frac{{\cal U}_0}{f_0L}\,,
\quad
\sigma = \frac{B_0}{L}\,,
\quad
{\cal F} = \frac{f_0^2 L^2}{gB_0}\,,
\label{Rossby-Froude}
\end{equation}
corresponding respectively to {\bfi Rossby number}, {\bfi aspect ratio}
and {\bfi (squared) rotational Froude number}. Typically, the Rossby
number and the aspect ratio are small, $\epsilon,\sigma \ll1$, while
the rotational Froude number is of order unity in atmospheric and
oceanic dynamics for $L$ at synoptic scales and larger. The {\bfi
nondimensional Euler fluid equations} corresponding to the Lagrangian
$l$ in equation (\ref{lag-Eul}) are obtained from the
Euler--Poincar\'e equations (\ref{EP-comp1}) or (\ref{EP-comp2}) with
$\mathbf{u} \rightarrow \mathbf{u}_3$ and $\nabla \rightarrow
\nabla_3$. Namely,
\begin{equation}
\epsilon\frac{d\mathbf{v}_3}{dt} - \mathbf{u}
\times {\rm curl}\,\mathbf{R}
+ \frac{1}{\epsilon{\cal F}}\hat{\bf z}
+ \frac{1}{(1+b)}\nabla_3\, p = 0\,.
\label{Eul-mot-nd}
\end{equation}
Clearly, the leading order balances in these equations are
hydrostatic in the vertical and geostrophic in the horizontal
direction. In this notation, the Kelvin--Noether circulation theorem
(\ref{KN-theorem-bD}) for the Euler fluid becomes
\begin{equation}\label{KN-theorem-Euler}
\frac{d}{dt}\oint_{\gamma_t(\mathbf{u}_3)}
(\epsilon\mathbf{v}_3+\mathbf{R})\cdot d\mathbf{x}_3
= -\oint_{\gamma_t(\mathbf{u}_3)}
\frac{1}{\rho_0(1+b)}\nabla_3 p\cdot d\mathbf{x}_3\;,
\end{equation}
Likewise, conservation of nondimensional potential
vorticity $q_{\rm Eul}$ on fluid parcels is given by, cf. equation
(\ref{pv-cons-Eul}),
\begin{equation} \label{pv-cons-Euler}
\frac{\partial{q}_{\rm Eul}}{\partial{t}}
+\mathbf{u}_3\cdot\nabla_3{q}_{\rm Eul} = 0\,,
\quad \hbox{where}\quad
q_{\rm Eul}=\nabla_3{b}\cdot\nabla_3
\times(\epsilon\mathbf{v}_3+\mathbf{R})\,.
\end{equation}

\paragraph{Hamilton's Principle Asymptotics.}
For sufficiently small buoyancy, $b=o(\epsilon)$, we define
$$
p\,' = {p}+\frac{z}{\epsilon{\cal F}}
\quad \hbox{and} \quad
b\,' =  \frac{b}{\epsilon{\cal F}}\,,
$$
and expand the Lagrangian (\ref{lag-Eul}) in powers of
$\epsilon$ as
\begin{equation}
l_{\rm EB} = \int dt\int d^{\,3}x\ D
\Big(\frac{\epsilon}{2}\mathbf{u}_3\cdot\mathbf{v}_3 +
\mathbf{u}\cdot\mathbf{R}(\mathbf{x})
 - b\,'z\Big)
 - p'(D-1) + o(\epsilon)\,.
\label{lag-v1}
\end{equation}
Upon dropping the order $o(\epsilon)$ term in the Lagrangian $l_{\rm
EB}$ the corresponding Euler--Poincar\'e equation gives the {\bfi
Euler-Boussinesq equation} for fluid motion in three dimensions,
namely,
\begin{equation}
\epsilon\frac{d\mathbf{v}_3}{dt}
- \mathbf{u} \times {\rm curl}\,\mathbf{R}
+ b\,'\hat{\bf z} + \nabla_3\, p\,' = 0\,,
\label{EB-mot}
\end{equation}
or, in horizontal and vertical components, with
${\rm curl}\,\mathbf{R}=f(\mathbf{x})\hat{\mbox{\boldmath{$z$}}}$,
\begin{eqnarray} \label{h+v.EB}
\epsilon\,\frac{d\mathbf{u}}{dt}
+ f\hat{\mbox{\boldmath{$z$}}}\times \mathbf{u}
+ \nabla\, p\,' = 0\,,
&&
\epsilon\, \sigma^2\,\frac{dw}{dt}+ b\,' +
\frac{\partial{p\,'}}{\partial{z}}   = 0\,,
\end{eqnarray}
where
\[
\frac{db\,'}{dt}=0\, \quad  \mbox{and} \quad
\nabla_3\cdot\mathbf{u}_3=\nabla\cdot\mathbf{u}
+\frac{\partial{w}}{\partial{z}}=0
\,.
\]
Even for order $O(\epsilon)$ buoyancy, the leading order balances are
still hydrostatic in the vertical, and geostrophic in the horizontal.
Equations (\ref{h+v.EB}) describe the motion of an ideal
incompressible stratified fluid relative to a stable
hydrostatic equlibrium in which the density is taken to be
constant except in the buoyant force. See, for example, Phillips
[1969] for a derivation of this approximate system via direct
asymptotic expansions of the Euler equations.
%
%

\paragraph{The Kelvin--Noether Theorem.} The Kelvin--Noether
circulation theorem (\ref{KN-theorem-bD}) for the Euler-Boussinesq
motion equation (\ref{EB-mot}) for an ideal incompressible stratified
fluid in three dimensions is,
\begin{equation}\label{KN-theorem-EB}
\frac{d}{dt}\oint_{\gamma_t(\mathbf{u}_3)}
(\epsilon\mathbf{v}_3+\mathbf{R})\cdot d\mathbf{x}
= -\oint_{\gamma_t(\mathbf{u}_3)} b\,' dz\;,
\end{equation}
where the curve $\gamma_t(\mathbf{u}_3)$ moves with the fluid
velocity $\mathbf{u}_3$. (The two Kelvin theorems in equations
(\ref{KN-theorem-Euler}) and (\ref{KN-theorem-EB})  differ in their
right hand sides.) By Stokes' theorem, the Euler-Boussinesq
equations generate circulation of  $\epsilon\mathbf{v}_3+\mathbf{R}$
around $\gamma_t(\mathbf{u}_3)$ whenever the gradient of bouyancy is
not vertical. Conservation of potential vorticity $q_{\rm EB}$ on
fluid parcels for the Euler-Boussinesq equations is given by
\begin{equation} \label{pv-cons-EB}
\frac{\partial q_{\rm EB}}{\partial{t}}+\mathbf{u}_3
\cdot\nabla_3\,q_{\rm
EB} = 0\,,
\quad \hbox{where}\quad
q_{\rm
EB}=\nabla_3{b\,'}\cdot\nabla_3\times(\epsilon\mathbf{v}_3
+\mathbf{R})\,.
\end{equation}

\subsection{Primitive Equations} \label{sec-PE}

\paragraph{The Lagrangian.} The primitive equations (PE) arise from the
Euler Boussinesq equations, upon imposing the approximation of
hydrostatic pressure balance. Setting the aspect ratio parameter
$\sigma$ to zero in the Lagrangian $l_{\rm EB}$ in equation
(\ref{lag-v1}) (see equations (\ref{uv-def}) and
(\ref{Rossby-Froude})), provides the Lagrangian for the nondimensional
primitive equations (PE),
\begin{equation}\label{lagPE}
l_{\rm PE} =  \int dt \int d^3 x
\left[ \frac{\epsilon}{2} D |\mathbf{u}|^2
+ D\mathbf{u}\cdot\mathbf{R}(\mathbf{x}) -
D b\,' z - p\,'(D-1)\right]\,.
\end{equation}
The Euler--Poincar\'e equations for $l_{\rm PE}$ now produce the PE;
namely, equations (\ref{h+v.EB}) with $\sigma=0$,
\begin{equation} \label{h+v.PE}
\epsilon\,\frac{d\mathbf{u}}{dt}
+ f\hat{\mbox{\boldmath{$z$}}}\times \mathbf{u}
+ \nabla\, p\,' = 0\,,
 b\,' +
\frac{\partial{p\,'}}{\partial{z}}   = 0\,,
\end{equation}
where
\[
 \frac{db\,'}{dt}=0\, \quad  \mbox{and} \quad
\nabla_3\cdot\mathbf{u}_3=\nabla\cdot\mathbf{u}
+\frac{\partial{w}}{\partial{z}}=0\,.
\]
Thus, from the
viewpoint of Hamilton's principle, imposition of hydrostatic balance
corresponds to ignoring the kinetic energy of vertical motion  by
setting $\sigma = 0$ in the nondimensional EB Lagrangian
(\ref{lag-v1}).

\paragraph{The Kelvin--Noether Theorem.} The Kelvin--Noether
circulation theorem for the primitive equations is obtained from
equation (\ref{KN-theorem-EB}) for the Euler-Boussinesq equations
simply by setting
$\sigma=0$. Namely,
\begin{equation}
\frac{d}{dt}\oint_{\gamma_t(\mathbf{u}_3)} (\epsilon\mathbf{u}
+\mathbf{R})\cdot d\mathbf{x}
= -\oint_{\gamma_t(\mathbf{u}_3)} b\,' dz\;,
\label{KN-theorem-Pe}
\end{equation}
where the curve $\gamma_t(\mathbf{u}_3)$ moves with the fluid
velocity $\mathbf{u}_3$. By Stokes' theorem, the primitive equations
generate circulation of $\epsilon\mathbf{u}+\mathbf{R}$ around
$\gamma_t(\mathbf{u}_3)$ whenever the gradient of bouyancy is not
vertical. The conservation of potential vorticity on fluid parcels
for the primitive equations is given by, cf. equation
(\ref{pv-cons-EP}),
\begin{equation} \label{pv-cons-PE}
\frac{\partial q_{\rm PE}}{\partial{t}}
+\mathbf{u}_3\cdot\nabla_3\,q_{\rm PE} = 0\,,
\quad \hbox{where}\quad
q_{\rm PE}=\nabla_3{b\,'}\cdot\nabla_3
\times(\epsilon\mathbf{u}+\mathbf{R})\,.
\end{equation}

\noindent{\bf Remark.}
In the limit, $\epsilon \rightarrow 0$,
Hamilton's principle for either
$l_{\rm EB}$, or $l_{\rm PE}$ gives,
\begin{equation}
f \hat{\mbox{\boldmath{$z$}}}\times\mathbf{u}
+ b\,'\,\hat{\mbox{\boldmath{$z$}}} + \nabla_3\, p\,' = 0,
\label{2bl}
\end{equation}
which encodes the leading order hydrostatic and
geostrophic equilibrium balances. These balances form the basis for
further approximations for near-geostrophic, hydrostatic flow.

\subsection{Hamiltonian Balance Equations} \label{sec-HBE}

\paragraph{Balanced Fluid Motions.} A fluid motion equation is said to
be {\bfi balanced}, if specification of the fluid's stratified buoyancy
and divergenceless velocity determines its pressure through the
solution of an equation which does not contain partial
time-derivatives among its highest derivatives. This definition of
balance makes pressure a diagnostic variable (as opposed to the
dynamic, or prognostic variables such as the horizontal velocity
components). The Euler equations (\ref{Eul-mot-nd}) and the
Euler-Boussinesq equations (\ref{EB-mot}) for the incompressible
motion of a rotating continuously stratified fluid are balanced in
this sense, because the pressure in these cases is determined
diagnostically from the buoyancy and velocity of the fluid by solving
a Neumann problem. However, the hydrostatic approximation of this
motion by the primitive equations (PE) is not balanced, because the
Poisson equation for the pressure in PE involves the time-derivative
of the horizontal velocity divergence, which alters the
character of the Euler system from which PE is derived and may lead to
rapid time dependence, as discussed in Browning et al. [1990].
Balanced approximations which eliminate this potentially rapid time
dependence have been sought and found, usually by using asymptotic
expansions of the solutions of the PE in powers of the small Rossby
number, $\epsilon\ll1$, after decomposing the horizontal velocity
$\mathbf{u}$ into order
$O(1)$ rotational and order $O(\epsilon)$ divergent components, as
$\mathbf{u}=\hat{\mbox{\boldmath{$z$}}}\times\nabla\psi+\epsilon\nabla\chi$,
where $\psi$ and $\chi$ are the stream function and velocity
potential, respectively, for the horizontal motion. (This is just
the Helmholtz decomposition with relative weight $\epsilon$.)

Balance equations (BE) are reviewed in the classic paper of
McWilliams and Gent [1980]. Succeeding investigations have concerned
the well-posedness and other features of various BE models
describing continuously stratified oceanic and atmospheric motions.
For example, consistent initial boundary value problems and regimes
of validity for BE are determined in Gent and McWilliams [1983a,b].
In other papers by these authors and their collaborators listed in
the bibliography, balanced models in isentropic coordinates are
derived, methods for the numerical solution of BE are developed, and
the applications of BE to problems of vortex motion on a
$\beta$-plane and wind-driven ocean circulation are discussed. In
studies of continuously stratified incompressible fluids, solutions
of balance equations that retain terms of order $O(1)$ and order
$O(\epsilon)$ in a Rossby number expansion of the PE solutions have
been found to compare remarkably well with numerical simulations of
the PE; see Allen, Barth, and Newberger [1990a,b] and Barth et al.
[1990]. Discussions of the relation between BE and semigeostrophy
have also recently appeared, see, e.g., Gent, McWilliams and Snyder
[1994] and Xu [1994].

\paragraph{Conservation of Energy and Potential Vorticity.} One
recurring issue in the literature is that, when truncated at order
$O(\epsilon)$ in the Rossby number expansion, the BE for continuously
stratified fluids conserve energy (Lorenz [1960]), but do not conserve
potential vorticity on fluid parcels.  Recently, Allen [1991] found a
set of BE for continuously stratified fluids that retains additional
terms of order
$O(\epsilon^2)$ and {\it does conserve potential vorticity} on fluid
parcels. Allen calls these balance equations ``BEM equations", because
they are based on momentum equations, rather than on the equation for
vertical vorticity, as for the standard BE. An advantage of the
momentum formulation of BEM over the vorticity formulation of the
original BE is that boundary conditions are more naturally imposed on
the fluid's velocity than on its vorticity. Holm [1996] derives
Hamiltonian balance equations (HBE) in the momentum formulation by
using the $\epsilon$-weighted Helmholtz decomposition for $\mathbf{u}$
and expanding Hamilton's principle (HP) for the PE in powers of the
Rossby number,
$\epsilon\ll1$. This expansion is truncated at order $O(\epsilon)$,
then all terms are retained that result from taking variations. As
we have seen, an asymptotic expansion of HP for the Euler-Boussinesq
(EB) equations which govern rotating stratified incompressible
inviscid fluid flow has two small dimensionless parameters: the
aspect ratio of the shallow domain, $\sigma$, and the Rossby number,
$\epsilon$. Setting $\sigma$ equal to zero in this expansion yields
HP for PE. Setting $\epsilon$ also equal to zero yields HP for
equilibrium solutions in both geostrophic and hydrostatic balance.
Setting $\sigma=0$, substituting the $\epsilon$-weighted Helmholtz
decomposition for $\mathbf{u}$ and truncating the resulting
asymptotic expansion in $\epsilon$ of the HP for the EB equations,
yields HP for a set  of nearly-geostrophic Hamiltonian balance
equations (HBE). The resulting HBE are equivalent to the BEM
equations in Allen [1991].

\paragraph{The Lagrangian.} The Lagrangian for the HBE model is given
in Holm [1996], cf. equation (\ref{lagPE}) for the PE action,
\begin{equation} \label{HBE-action}
\mathfrak{S}_{\rm HBE} = \int dt \int d^3 x \,
\Big[ D \mathbf{u}\cdot\mathbf{R}(\mathbf{x}) - D b z - p(D-1)
+\, \epsilon\frac{D}{2}
|\mathbf{u} - \epsilon\mathbf{u}_D|^2\Big]\,,
\end{equation}
where the horizontal fluid velocity is taken in balance equation form
as $\mathbf{u} = \mathbf{u}_R+\epsilon\mathbf{u}_D
= \hat{\mbox{\boldmath{$z$}}}\times\nabla\psi+\epsilon\nabla\chi$.
The corresponding Euler--Poincar\'e equations give the dynamics of the
HBE model
\begin{eqnarray}\label{hv-hbe0}
\epsilon \frac{d}{dt}\mathbf{u}_R+\epsilon^2 u_{Rj}\nabla u^j_D
+ f\hat{\mbox{\boldmath{$z$}}}\times\mathbf{u} + \nabla p &=& 0,
\nonumber \\
b + \frac{\partial{p}}{\partial{z}} + \epsilon^2 \mathbf{u}_R\cdot
\frac{\partial\mathbf{u}_{D}}{\partial{z}} &=& 0,
\nonumber \\
\hbox{with}\quad
\frac{db}{dt}=
\frac{\partial}{\partial{t}}\,b + \mathbf{u}\cdot\nabla{b}
+ \epsilon\, w \frac{\partial{b}}{\partial{z}} &=& 0,
\nonumber \\
\hbox{and}\quad
\nabla\cdot \mathbf{u}
+\epsilon\frac{\partial{w}}{\partial{z}}&=& 0.
\end{eqnarray}
Here the notation is the same as for the PE, except that
$w\rightarrow\epsilon\,w$ for HBE.

Dropping all terms of order $O(\epsilon^2)$ from the HBE model
equations (\ref{hv-hbe0}) recovers the BE discussed in Gent and
McWilliams [1983a,b]. Retaining these order $O(\epsilon^2)$ terms
restores the conservation laws due to symmetries of HP at the
truncation order $O(\epsilon)$. As explained in Holm [1996], the
resulting HBE model has the same order $O(\epsilon)$ accuracy as the
BE, since not {\it all} of the possible order $O(\epsilon^2)$ terms
are retained. Since the HBE model shares the same conservation laws and
Euler--Poincar\'e structure as EB and PE, and differs from them only
at order $O(\epsilon^2)$, it may be a valid approximation for times
longer than the expected order $O(1/\epsilon)$ for BE.

\paragraph{The Kelvin-Noether Theorem.} The HBE model (\ref{hv-hbe0})
possesses the following Kelvin--Noether circulation theorem,
\begin{equation} \label{kel-HBE}
\frac{d}{dt} \oint_{\gamma_t(\mathbf{u}_3)}
(\mathbf{R} + \epsilon\mathbf{u}_R) \cdot d\mathbf{x}_3
= - \oint_{\gamma_t(\mathbf{u}_3)} b dz,
\end{equation}
for any closed curve $\gamma_t(\mathbf{u}_3)$ moving with the fluid
velocity $\mathbf{u}_3$. We compare this result with the
Kelvin--Noether circulation theorem for PE in equation
(\ref{KN-theorem-Pe}), rewritten as
\begin{equation} \label{KN-Pe2}
\frac{d}{dt} \oint_{\gamma_t(\mathbf{u}_3)}
\underbrace{ (\mathbf{R} + \epsilon\mathbf{u}) }_{\hbox{PE}}
\cdot d\mathbf{x}_3 =
\frac{d}{dt} \oint_{\gamma_t(\mathbf{u}_3)}
\underbrace{ (\mathbf{R} + \epsilon\mathbf{u}_R }_{\hbox{HBE}}
+
\underbrace{\epsilon^2\mathbf{u}_D)
\cdot d\mathbf{x}_3 }_{\hbox{ZERO}}
= - \oint_{\gamma_t(\mathbf{u}_3)} b dz.
\end{equation}
Because
\[
\oint\mathbf{u}_D\cdot d\mathbf{x}_3=\oint d\chi=0,
\]
the $\epsilon^2$ term vanishes, and so the HBE circulation integral
differs from that of PE only through the differences in buoyancy
between the two theories.

The conservation of potential vorticity on
fluid parcels for the HBE model is given by, cf. equation
(\ref{pv-cons-EP}),
\begin{equation} \label{pv-cons-HBE}
\frac{\partial q_{\rm HBE}}{\partial{t}}
+\mathbf{u}_3\cdot\nabla_3\,q_{\rm HBE} = 0\,,
\quad \hbox{where}\quad
q_{\rm HBE}=\nabla_3\,{b}\cdot\nabla_3
\times(\epsilon\mathbf{u}_R+\mathbf{R})\,.
\end{equation}
Combining this with advection of $b$ and the tangential boundary
coundition on $\mathbf{u}_3$ yields an infinity of conserved
quantities,
\begin{equation} \label{csmrs1}
C_{\Phi}=\int d^3x  \ \Phi(q_{\rm HBE},b),
\end{equation}
for any function $\Phi$. These are the Casimir functions for the
Lie-Poisson Hamiltonian formulation of the HBE given in Holm [1996].

\paragraph{HBE Discussion.}
By their construction as Euler--Poincar\'e equations from a
Lagrangian which possesses the classic fluid symmetries, the
HBE conserve integrated energy and conserve potential vorticity on
fluid parcels. Their Lie-Poisson Hamiltonian structure endows the HBE
with the same type of self-consistency that the PE possess (for the
same Hamiltonian reason). After all, the conservation laws in both HBE
and PE are not accidental. They correspond to symmetries of the
Hamiltonian or Lagrangian for the fluid motion under continuous group
transformations, in accordance with Noether's theorem. In
particular, energy is conserved because the Hamiltonian in both
theories does not depend on time explicitly, and potential vorticity
is conserved on fluid parcels  because the corresponding Hamiltonian
or Lagrangian is right invariant under the infinite set of
transformations that relabel the fluid parcels without changing the
Eulerian velocity and buoyancy. See, e.g., Salmon [1988] for a review
of these ideas in the GFD context, as well as Holm, Marsden and Ratiu
[1998a,b] and the earlier sections of the present paper for the general
context for such results.

The vector fields which generate these relabeling transformations
turn out to be the {\it steady flows} of the HBE and PE models. By
definition, these steady flows leave invariant the Eulerian velocity
and buoyancy as they move the Lagrangian fluid parcels along the
flow lines. Hence, as a direct consequence of their shared
Hamiltonian structure, the steady flows of both HBE and PE are
relative equilibria. That is, steady HBE and PE flows are critical
points of a sum of conserved quantities, including the (constrained)
Hamiltonian. This shared critical-point property enables one, for
example, to use the Lyapunov method to investigate stability of
relative equilibrium solutions of HBE and PE. See Holm and Long
[1989] for an application of the Lyapunov method in the Hamiltonian
framework to the stability of PE relative equilibria. According to
the Lyapunov method, convexity of the constrained Hamiltonian at its
critical point (the relative equilibrium) is sufficient to provide a
norm that limits the departure of the solution from equilibrium
under perturbations. See, e.g., Abarbanel et al. [1986] for
applications of this method to the Euler equations for
incompressible fluid dynamics and Holm et al. [1985] for a range of
other applications in fluid and plasma theories.

Thus, the HBE arise as Euler--Poincar\'e equations and possess the
same Lie-Poisson Hamiltonian structure as EB and PE, and differ in
their Hamiltonian and conservation laws by small terms of order
$O(\epsilon^2)$. Moreover, the HBE conservation laws are
fundamentally of the same nature as those of the EB equations and
the PE from which they descend. These conserved quantities ---
particularly the quadratic conserved quantities --- may eventually
be useful measures of the deviations of the HBE solutions from EB
and PE solutions under time evolution starting from identical initial
conditions.

\subsection{Remarks on Two-dimensional Fluid Models in GFD}

The search for simpler dynamics than those of the primitive equations
natually leads to considerations of two-dimensional fluid models.
This certainly holds for applications in GFD, because the aspect
ratio of the domain ($\sigma$) and the Rossby number ($\epsilon$) of
the rotating flow are often small in these applications. Many
treatments of two-dimensional GFD models have been given using
asymptotic expansion methods in Hamilton's principle, see, e.g.,
Salmon [1983, 1985, 1988]. These treatments tend to focus especially on
the rotating shallow water (RSW) equations, their quasigeostrophic
(QG) approximation, and certain intermediate approximations, such as
the semigeostrophic (SG) equations (Eliassen [1949], Hoskins [1975],
Cullen and Purser [1989], Holm, Lifschitz and Norbury [1998]) and
the Salmon [1985] $L_1$ model.  A unified derivation of the RSW,
$L_1$, QG and SG equations using Hamilton's principle
asymptotics is given in Allen and Holm [1996]. This paper
also derives as Euler--Poincar\'e equations a new class of ``extended
geostrophic" (EG) models. The EG models include nonlocal corrections
to the ageostrophic velocity which could produce more accurate models
than the $L_1$, QG and SG approximations of the RSW equations.

There are also three dimensional versions of the QG and SG
equations, and recently a continuously stratified $L_1$ model was
derived in Allen, Holm and Newberger [1998] through the use of
Hamilton's principle asymptotics and the Euler--Poincar\'e theory.
For the suite of idealized, oceanographic, moderate Rossby number,
mesoscale flow test problems in Allen and Newberger [1993], this
continuously stratified $L_1$ model produces generally accurate
approximate solutions. These solutions are not quite as accurate as
those from the BEM/HBE or BE models, but are substantially more
accurate than those from three dimensional SG or QG.

Due to their wide applicability in GFD, the properties
of the two dimensional QG equations have been studied extensively.
Weinstein [1983] wrote down a Lie-Poisson bracket for them in
preparation for studying stability of quasigeostrophic equilibria.
The Hamiltonian structure and nonlinear stability of the equilibrium
solutions for the QG system and its variants has been thoroughly
explored. For references, see Marsden and Weinstein [1982],
Weinstein [1983] and Holm et al. [1985]. See also the introduction
and bibliography of Marsden et al. [1983] for a guide to
some of the  history and literature of this subject. A
discussion of the geodesic properties of the QG equations in the
framework of Euler--Poincar\'e theory is given in Holm, Marsden and
Ratiu [1998a,b]. A related discussion of QG in both two and three
dimensions is given from the viewpoint of Hamilton's principle
asymptotics in Holm and Zeitlin [1997].

Formulae showing the asymptotic expansion relationships among the
Lagrangians for the various GFD models are summarized in Tables (4.1)
and (4.2). In the next two sections, we turn our attention to dealing
with the mean effects of rapid fluctuations in GFD.

\newpage

\begin{table} \label{GFD approx}
\noindent
{\caption{\bf Table 1. Successive GFD approximations in Hamilton's
principle.}}
\begin{eqnarray}
l_{\rm Euler} &=& \int d^{\,3}x\,\Bigg[ D (1+b)
\Bigg(
\underbrace{ {\bf R}({\bf x})\cdot\mathbf{u} }_{\hbox{Rotation}}
 +
\underbrace{\frac{\epsilon}{2}|\mathbf{u}|^2
+ \frac{\epsilon}{2} \sigma^2 w^2 }_{\hbox{Kinetic
Energy}}\Bigg)
\nonumber \\
&& \hspace{1.5in}
- \underbrace{  D (1+b)
\Bigg(
\frac{z}{\epsilon{\cal F}}\Bigg) }_{\hbox{Potential Energy}}
 -
\underbrace{ p(D-1)  }_{\hbox{Constraint}}\Bigg]
\nonumber
\end{eqnarray}
\begin{description}
\item $\bullet \quad$
$l_{\rm Euler} \rightarrow l_{\rm EB}$, for small buoyancy,
$b=O(\epsilon)$.

\item $\bullet \quad$
$l_{\rm EB} \rightarrow l_{\rm PE}$, for small aspect ratio,
$\sigma^2=O(\epsilon)$.

\item $\bullet \quad$
$l_{\rm PE} \rightarrow l_{\frac{\rm HBE}{\rm BEM}}$, for
horizontal velocity
decomposition,
$\mathbf{u} = \hat{\mbox{\boldmath{$z$}}}\times \nabla \psi
+ \epsilon \nabla \chi
= \mathbf{u}_R + \epsilon \mathbf{u}_D $,
and $|\mathbf{u}|^2\rightarrow|\mathbf{u}_R|^2$ in $l_{\rm PE}$.

\item $\bullet \quad$
$l_{\frac{\rm HBE}{\rm BEM}} \rightarrow l_{\rm EG}$, upon rearranging KE in
$l_{\frac{\rm HBE}{\rm BEM}}$ and decomposing horizontal velocity as
$\mathbf{u} = \mathbf{u}_1 + \epsilon \mathbf{u}_2$, where
$\mathbf{u}_1 = \hat{\mbox{\boldmath{$z$}}}\times \nabla {\tilde\phi}$
with $${\tilde\phi}( {\bf x}_3,t) = \phi_S(x,y,t)
+\int^0_z dz'\ b\,,
$$
i.e., $\partial{\tilde\phi}/{ \partial z}=-\,b$ and
where $\mathbf{u}_2$ is the prescribed function,
\begin{equation} \label{agv1}
\mathbf{u}_2=\tau(\mathbf{u}_1\cdot\nabla)
\hat{\mbox{\boldmath{$z$}}}\times\mathbf{u}_1
-\,{\tilde\alpha}\tau\nabla\left(({\cal F}-\nabla^2)^{-1}
J({\tilde\phi},\psi)\right)    -\beta \tau f_1\mathbf{u}_1,
\nonumber
\end{equation}
with $\psi=f_1-b_1+\nabla^2 {\tilde\phi}$. The constants
$\tau$, ${\tilde\alpha}$, $\beta$, and $\gamma$ are free parameters and the
functions $f_1$ and $b_1$ denote prescribed order $O(\epsilon)$
Coriolis and topography variations.

\item $\bullet \quad$
$l_{\rm EG} \rightarrow l_{1}$, for horizontal velocity decomposition,
$\mathbf{u} = \mathbf{u}_1
=\hat{\mbox{\boldmath{$z$}}}\times \nabla {\tilde\phi}$ and dropping
terms of order $O(\epsilon^2)$ in $l_{\rm EG}$.

\item $\bullet \quad$
$l_{1} \rightarrow l_{QG}$, on dropping terms of order
$O(\epsilon^2)$ in the Euler--Poincar\'e equations for $l_{1}$.
\end{description}
\end{table}


\newpage

\begin{table} \label{EP-org}
\noindent
{\caption{\bf  Table 2. Nondimensional Euler--Poincar\'e Lagrangians
at successive levels of approximation via asymptotic expansions.}}
$$
l_{\rm Euler} = \int d^{\,3}x \left[ D (1+b)
\Bigg({\bf R}({\bf x})\cdot\mathbf{u}
 + \frac{\epsilon}{2}|\mathbf{u}|^2 + \frac{\epsilon}{2} \sigma^2 w^2
 - \frac{z}{\epsilon{\cal F}}\Bigg)
 - p(D-1)\right]
$$

$$
l_{\rm EB} = \int d^{\,3}x \left[ D
\Bigg({\bf R}\cdot\mathbf{u}
 + \frac{\epsilon}{2}|\mathbf{u}|^2 + \frac{\epsilon}{2} \sigma^2 w^2
 - b z \Bigg)
 - p(D-1)\right]
$$

$$
l_{\rm PE} = \int d^{\,3}x\left[ D
\Bigg({\bf R}\cdot\mathbf{u}
 + \frac{\epsilon}{2}|\mathbf{u}|^2
 - b z \Bigg)
 - p(D-1)\right]
$$

$$
l_{\frac{\rm HBE}{\rm BEM}} = \int d^{\,3}x\left[ D
\Bigg({\bf R}\cdot\mathbf{u}
 + \frac{\epsilon}{2}|\mathbf{u}- \epsilon \mathbf{u}_D|^2
 - b z\Bigg)
 - p(D-1)\right]
$$

$$
l_{\rm EG} = \int d^{\,3}x\left[ D
\Bigg(({\bf R} + \epsilon \mathbf{u}_1
 + \epsilon^2 \mathbf{u}_2 )\cdot\mathbf{u}
 - \frac{\epsilon}{2}|\mathbf{u}_1 + \gamma \epsilon \mathbf{u}_2 |^2
 - b z\Bigg)
 - p(D-1)\right]
$$

$$
l_{1} = \int d^{\,3}x\left[ D
\Bigg(({\bf R} + \epsilon \mathbf{u}_1)\cdot\mathbf{u}
 - \frac{\epsilon}{2}|\mathbf{u}_1 |^2
 - b z\Bigg)
 - p(D-1)\right]
$$

$$
l_{\rm QG/AW} =\!\! \int_{\mathcal{D}}
d^{\,2}x\!\!\int_{z_0}^{z_1}\!\! dz\left[ D
\Bigg(
\mathbf{R} \cdot\mathbf{u}
+{\frac{\epsilon}{2}}
\mathbf{u}\cdot(1-{\mathcal{L}}(z) {\Delta}^{-1})\mathbf{u}
\Bigg)
- p(D-1) \right],
$$
where
\[
{\mathcal{L}}(z) = \Big({\frac{\partial}{\partial
z}}+B\Big) {\frac{1}{{\mathcal{S}}(z)}}\Big({\frac{\partial}{\partial
z}}-B\Big) - {\mathcal{F}}
\]
and $B=0$ for QG.
\end{table}


\newpage

\section{Generalized Lagrangian Mean (GLM) Equations}
\label{sec-glmeqs}

The GLM theory of Andrews and McIntyre [1978a] is a hybrid
Eulerian-Lagrangian description in which Langrangian-mean flow
quantities satisfy equations expressed in Eulerian form. A related
set of equations was introduced by Craik and Leibovich [1976] in
their study of Langmuir vortices driven by a prescribed surface wave
field. The GLM equations are extended from prescribed fluctuation
properties to a theory of self-consistent Hamiltonian dynamics of
wave, mean-flow interaction for a rotating stratified incompressible
fluid in Gjaja and Holm [1996].

\paragraph{GLM Approximations.} In GLM theory, one decomposes the fluid
trajectory at fixed Lagrangian label
$\ell^A$, $A=1,2,3$, as follows,
\begin{equation}
\mathbf{x}^{\xi}(\ell^A,t) = \mathbf{x}(\epsilon \ell^A,\epsilon t) +
\alpha{\boldsymbol{\xi}}(\mathbf{x},t),
\quad\hbox{with}\quad\overline{{\boldsymbol{\xi}}}=0
\quad\hbox{and}\quad\overline{\mathbf{x}\cdot{\boldsymbol{\xi}}}=0\, ,
\label{X-dcmp}
\end{equation}
where scaling with $\epsilon$ denotes slow Lagrangian space and time
dependence, and an overbar denotes an appropriate
time average at fixed Eulerian position. For example, overbar may
denote the average over the rapid oscillation phase of a
single-frequency wave displacement ${\boldsymbol{\xi}}$ of amplitude
$\alpha$ relative to its wavelength. Thus, the displacement
${\boldsymbol{\xi}}(\mathbf{x},t)$ associated with such waves  has
zero Eulerian mean,
$\overline{{\boldsymbol{\xi}}(\mathbf{x},t)}=0$. Superscript $\xi$
on a function denotes its evaluation at the displaced Eulerian
position associated with the rapidly fluctuating component of the
fluid parcel displacement. Thus, e.g.,
$\mathbf{x}^{\xi} = \mathbf{x} + \alpha{\boldsymbol{\xi}}$ and
$\mathbf{u}^{\xi}(\mathbf{x}) = \mathbf{u}(\mathbf{x}^{\xi})$, for a function
$\mathbf{u}$.

The GLM operator $\overline{(\ )}^L$ averages over parcels at the
displaced positions
$\mathbf{x}^{\xi} = \mathbf{x} + \alpha{\boldsymbol{\xi}}$ and
produces slow Eulerian space and time dependence. This defines the
Lagrangian mean velocity:
\begin{equation}
\overline{\mathbf{u}}^L(\epsilon\mathbf{x},\epsilon{t})
\equiv\overline{\mathbf{u}(\mathbf{x},t)}^L
=\overline{\mathbf{u}^{\xi}(\mathbf{x},t)}
=\overline{\mathbf{u}(\mathbf{x}^{\xi},t)}\,,
\label{rbr}
\end{equation}
where unadorned overbar denotes the Eulerian average and scaling with
$\epsilon$ denotes slow dependence. Thus, the GLM description
associates to an instantaneous Eulerian velocity field
$\mathbf{u}(\mathbf{x}^{\xi},t)$ a unique Lagrangian mean velocity,
written (with a slight abuse of notation) as
$\overline{\mathbf{u}}^L(\epsilon\mathbf{x},\epsilon{t})$, such that
when a fluid parcel at  $\mathbf{x}^{\xi}=\mathbf{x} +
\alpha{\boldsymbol{\xi}}$  moves at its velocity
$\mathbf{u}(\mathbf{x}^{\xi},t)$, a fictional parcel at $\mathbf{x}$
is moving at velocity
$\overline{\mathbf{u}}^L(\epsilon\mathbf{x},\epsilon{t})$.  Hence,
\begin{equation}
\mathbf{u}(\mathbf{x} + \alpha{\boldsymbol{\xi}},t)
= \Big(\frac{\partial}{\partial t}\bigg|_\mathbf{x}
+ \overline{\mathbf{u}}^L\cdot{\boldsymbol{\nabla}} \Big)
\Big[ \mathbf{x}
+ \alpha{\boldsymbol{\xi}}(\mathbf{x},t) \Big]
\,=\, \overline{\mathbf{u}}^L
+ \alpha\overline{D}^L {\boldsymbol{\xi}}\,,
\label{stks}
\end{equation}
where $\overline{D}^L\equiv{\partial}/{\partial t}|_\mathbf{x} +
\overline{\mathbf{u}}^L\cdot{\boldsymbol{\nabla}}$ is the material
derivative with respect to the slowly varying Lagrangian mean
velocity, $\overline{\mathbf{u}}^L(\epsilon\mathbf{x},\epsilon{t})$.

In GLM theory one finds the following basic Eulerian relations and
definitions,
\begin{eqnarray}
\mathbf{u}^{\xi}(\mathbf{x},t) \equiv
\mathbf{u}(\mathbf{x}^{\xi},t) \!&=&\!
\overline{\mathbf{u}}^L(\epsilon
\mathbf{x},\epsilon t) + \alpha\mathbf{u}^l(\mathbf{x},t),
    \nonumber\\
\overline{\mathbf{u}}^L(\epsilon \mathbf{x},\epsilon
t)\!&\equiv&\!\overline{\mathbf{u}(\mathbf{x}
+\alpha{\boldsymbol{\xi}},t)},
    \nonumber\\
\mathbf{u}^l \equiv \overline{D}^L{\boldsymbol{\xi}}
\,,\!\!\!\!\!&&\!\!\!\!\!
\overline{D}^L\equiv \frac{\partial}{\partial{t}}\bigg|_\mathbf{x}
+\overline{\mathbf{u}}^L
\cdot{\boldsymbol{\nabla}}\,,
     \label{L_dcmp}\\
\mathbf{R}^{\xi}(\mathbf{x}) \equiv
\mathbf{R}(\mathbf{x}^{\xi}) \!&=&\!
\overline{\mathbf{R}}^L(\epsilon\mathbf{x})
+ \alpha\mathbf{R}^l({\boldsymbol{\xi}})\,,
    \nonumber\\
\overline{\mathbf{R}}^L(\epsilon\mathbf{x})
&\!\equiv&\!\overline{\mathbf{R}(\mathbf{x}
+\alpha{\boldsymbol{\xi}})},
\nonumber\\
\overline{\mathbf{u}^l}=\!\!&0&\!\!=\overline{\mathbf{R}^l}\,,
\nonumber
\end{eqnarray}
where $\mathbf{R}$ denotes the vector potential for the Coriolis
parameter, as before. The basic identity used in deriving these
formulae is
\begin{equation}
\Big(\frac{\partial}{\partial{t}}\bigg|_{\mathbf{x}^{\xi}}
+\mathbf{u}^{\xi}\cdot\frac{\partial}{\partial\mathbf{x}^{\xi}}
\Big)f(\mathbf{x}^{\xi},t)
=
\Big(\frac{\partial}{\partial{t}}\bigg|_\mathbf{x}
+\overline{\mathbf{u}}^L\cdot
\frac{\partial}{\partial\mathbf{x}}
\Big)f(\mathbf{x}+\alpha{\boldsymbol{\xi}},t),
\label{ttd}
\end{equation}
which holds for any differentiable function $f$. This identity may
be shown by taking the partial time derivative at constant $\ell^A$
of the decomposition (\ref{X-dcmp}) and using the chain rule, cf.
Andrews and McIntyre [1978a].
\medskip

\paragraph{The Lagrangian.} We return to Hamilton's principle with
Lagrangian (\ref{lag-v1}) for the Euler-Boussinesq equations. This is
expressed in terms of unscaled (i.e., dimensional) Eulerian
instantaneous quantities in the GLM notation as
\begin{eqnarray}
\mathfrak{S}_{\rm EB} &=& \int dt\int d^3 x
  \Bigg\{
 D^{\xi}
  \Bigg[ \frac{1}{2} |\mathbf{u}^{\xi}(\mathbf{x},t)|^2
  -b^{\xi}g z
+\ \mathbf{u}^{\xi}(\mathbf{x},t)
\cdot\mathbf{R}^{\xi}(\mathbf{x})\Bigg]
\nonumber \\
&&\hspace{1.5in}  - p^{\xi}\left(D^{\xi} - 1\right)
  \Bigg\},
\label{l}
\end{eqnarray}
where $g$ is the constant acceleration of gravity and
\begin{equation} \label{D-xi}
D^{\xi}(\mathbf{x},t) = D(\mathbf{x}+\alpha{\boldsymbol{\xi}},t)
= {\rm det}\Big(\delta^i_j
+ \alpha \frac{\partial\xi^i}{\partial{x^j}}
\Big)\,,
\end{equation}
which is a cubic expression in $\alpha$.
We denote the corresponding pressure decomposition as,
\begin{equation}
p^{\xi}(\mathbf{x},t)
= p(\mathbf{x}+\alpha{\boldsymbol{\xi}},t)
= \overline{p}^L(\epsilon\mathbf{x},\epsilon{t})
+ \sum_{j=1}^3 \alpha^j
h_j(\epsilon\mathbf{x},\epsilon{t})p_j(\mathbf{x},t)\,.
\label{pdcmp_lag}
\end{equation}
After expanding $D^{\xi}$ in powers of $\alpha$, Gjaja and Holm
[1996] average over the rapid space and time scales in the action
(\ref{l}) for the Euler-Boussinesq equations while keeping the
Lagrangian coordinates $\ell^A$ and $\epsilon t$ fixed. (This is the
Lagrangian mean of the action.)

\paragraph{Remark.} Note that averaging in this setting
is a formal operation associated with the addition of a new degree
of freedom describing the rapid fluctuations. Thus, averaging in
itself does not entail any approximations. The approximations occur
next, in the  truncations of the expansions of the averaged action
in powers of the small parameters $\epsilon$ and $\alpha$.\smallskip

The GLM dynamics follows upon making the decompositions in
(\ref{X-dcmp})-(\ref{stks}) and (\ref{D-xi})-(\ref{pdcmp_lag}),
averaging the action (\ref{l}) and assuming that the rapidly
fluctuating displacement
${\boldsymbol{\xi}}$ is a {\it prescribed} function of
$\mathbf{x}$ and $t$, which satisfies the transversality condition
given in (\ref{kdota_AM}) below.

The averaged, truncated action is
(taking $\mathbf{R}(\mathbf{x})={\boldsymbol{\Omega}}
\times\mathbf{x}$ with constant rotation frequency
${\boldsymbol{\Omega}}$ for simplicity)
\begin{eqnarray}
\overline{\mathfrak{S}}_{GLM} &=& \int dt\int d^3x\Bigg\{ D\Bigg[
  \frac{1}{2} |\overline{\mathbf{u}}^L|^2
+\frac{\alpha^2}{2}
\overline{\bigg|
\frac{\partial{\boldsymbol{\xi}}}{\partial t}
+ (\overline{\mathbf{u}}^L\cdot{\boldsymbol{\nabla}})
{\boldsymbol{\xi}}
\bigg|^2}
-b\Big(\ell^A(\mathbf{x},t)\Big)gz
\nonumber\\
&&+\,
\overline{\mathbf{u}}^L\cdot({\boldsymbol{\Omega}}\times\mathbf{x})
+\alpha^2\overline{\Big(
\frac{\partial{\boldsymbol{\xi}}}{\partial t}
+ (\overline{\mathbf{u}}^L
\cdot{\boldsymbol{\nabla}}){\boldsymbol{\xi}}\Big)
\cdot({\boldsymbol{\Omega}}\times{\boldsymbol{\xi}})}
\,\Bigg]
\label{lbar_AM}\\
&&
+\ \overline{p}^L\Bigg[1 - D
+\frac{\alpha^2}{2}\frac{\partial}{\partial x^i}
\overline{ \Big(\xi^i\frac{\partial\xi^j}{\partial x^j}
-
\xi^j\frac{\partial\xi^i}{\partial x^j}\Big)}\,\Bigg]
+\alpha^2 h_1 \overline{
\left(p_1\frac{\partial\xi^j}{\partial x^j}\right)}
\nonumber \\
&&+O(\alpha^4).
\Bigg\}
\nonumber
\end{eqnarray}
Note that the buoyancy $b$ is a function of the Lagrangian labels
$\ell^A$ which is held fixed during the averaging.

The variation of
$\overline{\mathfrak{S}}_{GLM}
=\int\,dt\overline{L}_{GLM}$ in
equation (\ref{lbar_AM}) with respect to
$h_1$ at fixed $\mathbf{x}$ and $t$
yields the transversality condition,
\begin{equation}
\overline{
\left(p_1\frac{\partial\xi^j}{\partial x^j}\right)}
= O(\alpha^2\epsilon).
\label{kdota_AM}
\end{equation}
When ${\boldsymbol{\xi}}$ and $p_1$ are single-frequency wave
oscillations, this condition implies that the wave amplitude is
transverse to the wave vector; hence the name ``transversality
condition.''  Gjaja and Holm [1996] show that this condition is
required for the Euler--Poincar\'e equations resulting from
averaging in Hamilton's principle to be consistent with applying the
method of averages directly to the Euler-Boussinesq equations.

Next, the variation of $\overline{\mathfrak{S}}_{GLM}$ with respect
to $\overline{p}^L$ at fixed $\mathbf{x}$ and $t$ gives
\begin{equation}
D = 1
+\frac{\alpha^2}{2}\frac{\partial}{\partial x^i}
\overline{\Big(
  \xi^i\frac{\partial\xi^j}{\partial x^j}
- \xi^j\frac{\partial\xi^i}{\partial x^j}\Big)},
\label{d1_AM}
\end{equation}
where the second term on the right side is order
$O(\alpha^2\epsilon)$, and thus is negligible at order
$O(\alpha^2)$. This follows because mean quantities only depend on
slow space and slow time, for  which
$\partial/{\partial x^j}=\epsilon\partial/\partial(\epsilon{x^j})$.

\paragraph{The Euler--Poincar\'e Equations.} The
Euler--Poincar\'e equation for the action
$\overline{\mathfrak{S}}_{GLM}$ results in the GLM motion equation,
\begin{eqnarray}
&&
  \frac{\partial({\bf{m}}/D)}{\partial\epsilon t} -
\overline{\mathbf{u}}^L\times
  \bigg(\frac{\partial}{\partial\epsilon\mathbf{x}}
  \times\frac{\bf {m}} {D}\bigg)
+\ \frac{1}{\epsilon} b g \hat{\mbox{\boldmath{$z$}}}
\label{meq_AM}\\
&&
+\
\frac{\partial}{\partial\epsilon\mathbf{x}}
\left[\overline{p}^L
+ |\overline{\mathbf{u}}^L|^2
- \frac{1}{2}\overline{\mathbf{u}^\xi\cdot\mathbf{u}^\xi}
- \overline{\mathbf{u}^\xi\cdot({\boldsymbol{\Omega}}
\times{\boldsymbol{\xi}})}
- \alpha^2{\bf{p}}\cdot\overline{\mathbf{u}}^L\right] =  0,
\nonumber
\end{eqnarray}
where the Lagrangian mean momentum $\mathbf{m}$ is defined as
\begin{equation} \label{m-def}
  {\bf{m}} \equiv \frac{\delta\overline{L}_{GLM}}
  {\delta\overline{\mathbf{u}}^L}
= D\big[\overline{\mathbf{u}}^L
+ ({\boldsymbol{\Omega}}\times\mathbf{x})\big]
-\alpha^2{\bf p},
\end{equation}
and (leaving $\alpha^2$ explicit) ${\bf p}$ is the ``pseudomomentum
density,'' defined by, cf. Andrews and McIntyre [1978a],
\begin{equation}
{\bf{p}} \equiv -\,D\ \overline{(u^l_j+R^l_j) {\boldsymbol{\nabla}}
\xi^j}
\quad\hbox{and}\quad
D\equiv\hbox{det}\bigg(\frac{\partial \ell^A}{\partial x^j}\bigg)\,.
\label{def-psmom}
\end{equation}
Equations (\ref{d1_AM}) and (\ref{meq_AM})
(and implicitly (\ref{kdota_AM}))
are the equations of the GLM theory for
incompressible flow discussed in
Andrews and McIntyre [1978a].

\paragraph{The Kelvin--Noether Theorem.} The Kelvin--Noether
circulation theorem for the GLM motion equation (\ref{meq_AM}) states
that
\begin{equation} \label{KN-theorem-GLM}
\frac{d}{d \epsilon t}
\oint_{\overline{\gamma}(\epsilon t)}
\mathbf{m}{\cdot}d\mathbf{x}
= -\,\frac{g}{\epsilon}
\oint_{\overline{\gamma}(\epsilon t)} b \, dz\,,
\end{equation}
where the curve ${\overline{\gamma}(\epsilon t)}$ moves with the
Lagrangian mean fluid velocity $\overline{\mathbf{u}}^L$. By Stokes'
theorem, the GLM equations generate circulation of $\mathbf{m}$ around
${\overline{\gamma}(\epsilon t)}$ whenever the gradient of bouyancy
is not vertical. The conservation of potential vorticity on fluid
parcels for the GLM equations is given by
\begin{equation} \label{pv-cons-GLM}
\overline{D}^L q_{GLM}=0\,,
\quad \hbox{where}\quad
q_{GLM}=\frac{1}{D}\frac{\partial b}{\partial\epsilon\mathbf{x}}
\cdot\left(\frac{\partial}{\partial\epsilon\mathbf{x}}
  \times\frac{\bf {m}} {D}\right)\,.
\end{equation}
The constraint $D=1$ is imposed at order $O(\alpha^2)$ in Hamilton's
principle for the action (\ref{lbar_AM}) by the slow component of
the pressure. This condition holds at order $O(\alpha^2)$ when
${\boldsymbol{\nabla}}_{\epsilon\mathbf{x}}\cdot\overline{\mathbf{u}}^L=0$,
since $D$ satisfies $\partial D/\partial\epsilon
t+{\boldsymbol{\nabla} }_{\epsilon\mathbf{x}}\cdot
D\overline{\mathbf{u}}^L=0$.

\paragraph{Alternative Derivation of the Kelvin--Noether Circulation
Theorem for GLM.} The {\it unapproximated} Euler-Boussinesq
equations (\ref{EB-mot}) for a rotating stratified incompressible
fluid are, in the GLM notation,
\begin{eqnarray}
\left(\frac{\partial}{\partial{t}}
\bigg|_{\mathbf{x}^{\xi}}+\mathbf{u}^{\xi}\cdot
\frac{\partial}{\partial\mathbf{x}^{\xi}}\right)\mathbf{u}^{\xi}
\,-\,\mathbf{u}^{\xi}\times 2{\boldsymbol{\Omega}}(\mathbf{x}^{\xi})
&=& -\,\frac{\partial{p}^{\xi}}{\partial\mathbf{x}^{\xi}}- b^{\xi} g
\hat{\mbox{\boldmath{$z$}}}^{\xi},
\label{Eul-eqs}\\
\left(\frac{\partial}{\partial{t}}\bigg|_{\mathbf{x}^{\xi}}
+\mathbf{u}^{\xi}\cdot
\frac{\partial}{\partial\mathbf{x}^{\xi}}\right)b^{\xi}
= 0,\quad
\frac{\partial}{\partial\mathbf{x}^{\xi}}\cdot\mathbf{u}^{\xi} = 0,
\!\!\!\!&&\!\!\!\!
2{\boldsymbol{\Omega}}(\mathbf{x}^{\xi})=
\frac{\partial}{\partial\mathbf{x}^{\xi}}
\times\mathbf{R}(\mathbf{x}^{\xi})\,.
\nonumber
\end{eqnarray}
Being Euler--Poincar\'e, these equations admit the following
Kelvin--Noether circulation theorem, cf.
equation (\ref{KN-theorem-EB}),
\begin{equation}\label{KN-theorem-EB-GLM}
\frac{d}{d t}\oint_{\gamma(t)}
\Big(\mathbf{u}(\mathbf{x}^{\xi},t)+\mathbf{R}(\mathbf{x}^{\xi})\Big)
\cdot d\mathbf{x}^{\xi}
= -\,g
\oint_{\gamma(t)} b(\mathbf{x}^{\xi},t)\ dz^{\xi},
\end{equation}
for a contour $\gamma(t)$ which moves with the fluid.
\medskip

The time average of the Kelvin circulation integral is expressed as
\begin{eqnarray}
\overline{I}(\epsilon t) &=&\overline{\oint_{\gamma(t)}
\Big(\mathbf{u}(\mathbf{x}^{\xi},t)
+\mathbf{R}(\mathbf{x}^{\xi})\Big)
\cdot d\mathbf{x}^{\xi}}
\nonumber \\
&=&
\overline{\oint_{\gamma(t)}
(\overline{\mathbf{u}}^L
+\overline{\mathbf{R}}^L+\alpha\mathbf{u}^l
+\alpha\mathbf{R}^l) \cdot
(d\mathbf{x}+\alpha\,d{\boldsymbol{\xi}})}
\nonumber\\
&=& \oint_{\overline{\gamma}(\epsilon t)}
\bigg[(\overline{\mathbf{u}}^L+\overline{\mathbf{R}}^L)\cdot d\mathbf{x}
+\alpha^2\overline{(\mathbf{u}^l+\mathbf{R}^l) \cdot
d{\boldsymbol{\xi}}}\,\bigg]
\,,
\label{kelnt-bar}
\end{eqnarray}
where the contour ${\overline{\gamma}(\epsilon t)}$ moves with
velocity $\overline{\mathbf{u}}^L$, since it follows the fluid
parcels as the average is taken. Using equations (\ref{m-def}) and
(\ref{def-psmom}), we rewrite
$\overline{I}(\epsilon t)$ as
\begin{equation}
\overline{I}(\epsilon t) = \oint_{\overline{\gamma}(\epsilon t)}
\Big(\overline{\mathbf{u}}^L+\overline{\mathbf{R}}^L
- \alpha^2{\bf{p}}/D\Big)\cdot
d\mathbf{x}
= \oint_{\overline{\gamma}(\epsilon t)}\mathbf{m}\cdot
d\mathbf{x}
\,,
\label{I-psmom}
\end{equation}
which re-introduces Lagrangian mean momentum
$\mathbf{m}(\epsilon\mathbf{x},\epsilon t)$ and the pseudomomentum
density of the rapid motion,
${\bf{p}}(\epsilon\mathbf{x},\epsilon t)$.
In terms of these quantities, we
may write the Lagrangian mean of the Kelvin--Noether
circulation theorem for
the Euler-Boussinesq equations as
\begin{equation} \label{avg-eb-kn}
\frac{d}{dt} \overline{I}(\epsilon t)
= \epsilon\frac{d}{d \epsilon t}
\oint_{\overline{\gamma}(\epsilon t)}\mathbf{m}\cdot
d\mathbf{x}
= -\,g \overline{\oint_{\gamma(t)} b(\mathbf{x}^{\xi},t)\ dz^{\xi}}
= -\,g \oint_{\overline{\gamma}(\epsilon t)} b \, dz\,.
\end{equation}
This result recovers the Kelvin--Noether circulation theorem
(\ref{KN-theorem-GLM}) for the GLM equations (\ref{meq_AM}) which
were derived above as Euler--Poincar\'e equations for the averaged
Lagrangian $\overline{\mathfrak{S}}_{GLM}$ in equation
(\ref{lbar_AM}).

\paragraph{Geometry of the Stokes Mean Drift Velocity and the
Pseudomomentum.} The quantity ${\bf{p}}$ in equation
(\ref{def-psmom}) is an additional slowly varying fluid degree of
freedom which emerges in the process of averaging to describe the
rectified effects of the rapidly varying component of the fluid
motion acting on the slowly varying component. The rectified effects
of wave fluctuations in fluids are traditionally discussed in terms
of another quantity, namely, the Stokes mean drift velocity, defined as
\begin{equation} \label{stokes-def}
\overline{\mathbf{u}}^S
= \overline{\mathbf{u}}^L - \overline{\mathbf{u}}
= \alpha^2\overline{({\boldsymbol{\xi}}
\cdot{\boldsymbol{\nabla}})\mathbf{u}^l} + O(\alpha^4)\,,
\end{equation}
We define the analogous Stokes mean drift quantity corresponding to
the rotation vector potential, $\mathbf{R}$, as
\begin{equation} \label{R-stokes-def}
\overline{\mathbf{R}}^S
= \overline{\mathbf{R}}^L - \overline{\mathbf{R}}
= \alpha^2\overline{({\boldsymbol{\xi}}
\cdot{\boldsymbol{\nabla}})\mathbf{R}^l} + O(\alpha^4)\,,
\end{equation}
The time-averaged contour integral
$\overline{I}(\epsilon t)$ in equation
(\ref{kelnt-bar}) may be rewritten as
\begin{eqnarray} \label{Kel}
\overline{I}(\epsilon t) &=&
 \oint_{\overline{\gamma}(\epsilon t)}
\bigg[(\overline{\mathbf{u}}+\overline{\mathbf{R}})\cdot d\mathbf{x}
+ (\overline{\mathbf{u}}^S
+\overline{\mathbf{R}}^S)\cdot d{\mathbf{x}}
- {\bf p}\cdot d{\mathbf{x}}\,\bigg]
\nonumber \\
&=&
 \oint_{\overline{\gamma}(\epsilon t)}
\bigg[(\overline{\mathbf{u}}
+\overline{\mathbf{R}})\cdot d\mathbf{x}
+ \alpha^2\overline{\bigg( {\boldsymbol{\xi}}\cdot
{\boldsymbol{\nabla}}
(\mathbf{u}^l+\mathbf{R}^l)\bigg)}\cdot d{\mathbf{x}}
+\alpha^2\overline{(\mathbf{u}^l+\mathbf{R}^l) \cdot
d{\boldsymbol{\xi}}}\,\bigg]
\nonumber \\
&=&  \oint_{\overline{\gamma}(\epsilon t)}
\bigg[(\overline{\mathbf{u}}+\overline{\mathbf{R}})\cdot d\mathbf{x}
+ \alpha^2\overline{\bigg(
\pounds_{\xi}
\Big( (\mathbf{u}^l+\mathbf{R}^l)\cdot
d{\mathbf{x}}\Big)\bigg)}\ \bigg]\,.
\end{eqnarray}
Thus, the last two terms combine into the time-averaged Lie
derivative with respect to ${\boldsymbol{\xi}}$ of the total
fluctuation circulation $(\mathbf{u}^l+\mathbf{R}^l)\cdot
d\mathbf{x}$. This is a remarkable formula which shows the geometric
roles of the Stokes mean drift velocity and the pseudomomentum: The
Stokes mean drift velocity derives from {\it transport}, while the
pseudomomentum is caused by {\it line-element stretching} by the
fluctuations. For {\it divergence-free} fluctuations, we have
${\boldsymbol{\nabla}}\cdot{\boldsymbol{\xi}}=0$, and the Stokes
mean drift velocity in the rotating frame is a higher order term. In
particular,
\begin{equation} \label{Stk-drft}
\alpha^2\bar{\mathbf{u}}^S
= \alpha^2\overline{\bigg(
({\boldsymbol{\xi}}\cdot{\boldsymbol{\nabla}})
\frac{d\boldsymbol{\xi}}{dt}\bigg)}
= \alpha^2\frac{\partial}{\partial{x^i}}\overline{\bigg({\xi}^i\
\frac{d\boldsymbol{\xi}}{dt}\bigg)}
= \epsilon\ \alpha^2\frac{\partial}{\partial\epsilon{x^i}}
\overline{\bigg({\xi}^i\
\frac{d\boldsymbol{\xi}}{dt}\bigg)}\,.
\end{equation}
The last step in this calculation follows because mean quantities
only depend on slow space and slow time. Thus, the Stokes mean drift
velocity (i.e., the difference between the Lagrangian and Eulerian
mean velocities) is order $O(\epsilon)$ smaller than the
pseudomomentum, for divergence-free fluctuations.

\section{Nonlinear Dispersive Modifications of the\\
 EB Equations and PE}
\label{sec-MEB}

In generalizing earlier work by Camassa and Holm [1993] from one
dimension to $n$ dimensions, Holm, Marsden and Ratiu [1998a,b]
used the Euler--Poincar\'e framework to formulate a modified set of
Euler equations for ideal homogeneous incompressible fluids. This
modification introduces nonlinear dispersion into Euler's equations,
which is designed physically to model {\it nondissipative} unresolved
rapid fluctuations. This nonlinear dispersion is founded
mathematically on the geometrical property that solutions of the basic
Euler--Poincar\'e equations are geodesics on an underlying group
when their Lagrangian is a metric. Here we use the
Euler--Poincar\'e reduction theorems (\ref{rarl}) and
(\ref{EPforcontinua}) including advected parameters and Hamilton's
principle asymptotics in the EB Lagrangian in GLM notation (\ref{l})
to formulate a modified set of Euler-Boussinesq equations that includes
nonlinear dispersion along with stratification and rotation. In this
new modification of the Euler-Boussinesq equations, nonlinear
dispersion adaptively filters high wavenumbers and thereby enhances
stability and regularity without compromising either low wavenumber
behavior, or geophysical balances. Here, `high' and `low' refer
repectively to wavenumbers greater, or less than the inverse of a
fundamental length scale $\alpha$, which parameterizes the nonlinearly
dispersion in the model. We also present the corresponding nonlinear
dispersive modification of the primitive equations.  We leave it as an
open question, whether our nonlinearly dispersive primitive equation
model will have a slow manifold when dissipation and forcing are
included.

\subsection{Higher Dimensional Camassa--Holm Equation.}

\paragraph{The Lagrangian and Action Functionals.} As shown in Holm,
Marsden and Ratiu [1998a,b], the Camassa-Holm (CH) equation (Camassa
and Holm [1993]) in $n$ dimensions describes geodesic motion on the
diffeomorphism group of
$\mathbb{R}^n$ with respect to the metric given by the $H^1$ norm of
the Eulerian fluid velocity. This Euler--Poincar\'e equation follows
from the Lagrangian
$l_{\rm CH}$ given by the $H^1$ norm of the fluid velocity
$\mathbf{u}$ in $n$ dimensions, subject to volume preservation (for
$n\ne1$), namely,
\begin{equation}\label{CH-lag}
\mathfrak{S}_{\rm CH}
= \int dt \; l_{\rm CH} = \int dt\int_{\cal M} d^{\,n}x\
\bigg\{\,\frac{D}{2} \Big(|\mathbf{u}|^2
+ \alpha^2 |\nabla\mathbf{u}|^2\Big) - p(D-1)\bigg\}\,,
\end{equation}
in which the parameter $\alpha$ has dimensions of length. We
denote
$({\nabla}\mathbf{u})^i_j=u^i_{,j}
  \equiv{\partial}u^i/{\partial}x^j$,
$|\nabla\mathbf{u}|^2\equiv u^i_{,j}u_i^{,j}= {\rm
tr}(\nabla\mathbf{u}\cdot\nabla\mathbf{u}^T)$,
tr is the trace operation
for matrices and the superscript $({ \cdot})^T$ denotes transpose.

The action $\mathfrak{S}_{\rm CH}$ in (\ref{CH-lag}) is the
order $O(\alpha^2)$ expression for the mean of the EB action in
equation (\ref{l}) in the absence of stratification and rotation,
namely
\begin{equation}
\mathfrak{S}_{\rm CH} \approx \int dt\int d^n x\
  \bigg\{ D
  \frac{1}{2}
\overline{|\mathbf{u}^{\xi}(\mathbf{x},t)|^2}
  - p\left(D - 1\right)
  \bigg\},
\label{l-ch}
\end{equation}
in which we neglect the order $O(\alpha^2\epsilon)$ pressure and
density corrections discussed in Gjaja and Holm [1996] and approximate
the mean kinetic energy in a Taylor expansion as follows,
\begin{eqnarray} \label{E-avg}
\frac{1}{2}\overline{|\mathbf{u}^{\xi}(\mathbf{x},t)|^2}
&=& \frac{1}{2}\overline{|\mathbf{u}(\mathbf{x} +
\alpha{\boldsymbol{\xi}},t)|^2}
\ =\ \frac{1}{2}\overline{|\mathbf{u}(\mathbf{x},t) +
\alpha{\boldsymbol{\xi}}\cdot{\boldsymbol{\nabla}}\mathbf{u}|^2}
+ O(\alpha^4)
\nonumber \\
&=&
  \frac{1}{2} |\mathbf{u}|^2 + \frac{\alpha^2}{2}
\overline{|
{\boldsymbol{\xi}}\cdot{\boldsymbol{\nabla}}\mathbf{u}|^2}
+ O(\alpha^4)
\ \approx\
  \frac{1}{2} |\mathbf{u}|^2 + \frac{\alpha^2}{2}
  |{\boldsymbol{\nabla}}\mathbf{u}|^2\,.\hspace{.25in}
\end{eqnarray}
In the last step, we have also dropped terms of order $O(\alpha^4)$
and assumed isotropy of the rapid fluctuations, so that
$\overline{\xi^i\xi^j}\approx\delta^{ij}$. As in GLM theory, the
length-scale parameter $\alpha$ represents the amplitude of the
rapidly fluctuating component of the fluid parcel trajectory (or its
amplitude to length-scale ratio in a nondimensional formulation).

\paragraph{The Euler--Poincar\'e Equations.} Varying the action
$\mathfrak{S}_{\rm CH}$ in (\ref{CH-lag}) at fixed
$\mathbf{x}$ and $t$ gives
\begin{eqnarray}
\delta \mathfrak{S}_{\rm CH}
&=&\!\! \int dt\int_{\cal M} d^{\,n}x \,\Big[
\left( \frac{1}{2}|\mathbf{u}|^2
+ \frac{\alpha^2}{2}|\nabla\mathbf{u}|^2-p\right) \delta D \nonumber
\\  &&\hspace{.3in}
+ \left(D \mathbf{u}
- \alpha^2 ({\rm div}D{\rm grad}) \mathbf{u} \right)\cdot
\delta \mathbf{u} -(D-1)\delta p\Big]
\nonumber \\
&&\hspace{.3in}
+\ \alpha^2 \int dt\oint_{\partial{\cal M}} d^{\,n-1}x \
(D\hat{\mbox{\boldmath{$n$}}}\cdot
\nabla \mathbf{u} \cdot \delta \mathbf{u})
\,,
\label{var.ndCH}
\end{eqnarray}
whose natural boundary conditions on $\partial{\cal M}$ are
\begin{equation}\label{boundary.equation}
\mathbf{u} \cdot \hat{\mathbf{n}} = 0
\quad {\rm and} \quad
(\hat{\mathbf{n}} \cdot \nabla)
\mathbf{u}\ \|\ \hat{\mathbf{n}},
\end{equation}
where $\|$ denotes ``parallel to" in the second boundary condition,
which of course is not imposed when $\alpha^2$ is absent. (Recall
that $\delta\mathbf{u}$ in equation (\ref{var.ndCH}) is arbitrary
except for being tangent on the boundary. This tangency, along with
the second condition in equation (\ref{boundary.equation}) is
sufficient for the boundary integral in equation (\ref{var.ndCH}) to
vanish.)

Unfortunately, the space of divergence free vector fields with the
boundary conditions (\ref{boundary.equation}) do not form a Lie
algebra unless the boundaries are flat. There are two ways to deal
with this problem. One is to use the no-slip boundary conditions $u =
0$ (and hence $\delta u = 0$ on $\partial \mathcal{M}$). This is
especially interesting in conjunction with the addition of viscosity.
Another is to modify these boundary conditions with the addition of a
second fundamental form term to the right hand side. The latter
idea is explored in detail in Holm, Kouranbaeva, Marsden,
Ratiu, and Shkoller [1998]. In what follows, we imagine that one of
these options is chosen.

By equation (\ref{EP-comp1}), the Euler--Poincar\'e equation for the
action $\mathfrak{S}_{\rm CH}$ in equation (\ref{CH-lag}) is
\begin{equation}\label{nd:CHeqn}
\left(\frac{ \partial}{\partial t}
+ \mathbf{u}\cdot\nabla\right)\mathbf{v}
+v_j\nabla u^j
+\nabla\left(p - \frac{1}{2}|\mathbf{u}|^2
- \frac{\alpha^2}{2}|\nabla\mathbf{u}|^2\right) = 0\,,
\end{equation}
where
\[
 \mathbf{v} \equiv \mathbf{u} - \alpha^2\Delta \mathbf{u}
= \frac{1}{D} \frac{{\delta} l_{\rm CH}}
{{\delta} \mathbf{u}}\Bigg|_{D=1}\,.
\]
In writing this equation, we have substituted the constraint $D=1$,
which as before implies incompressibility via the continuity equation
for $D$. Requiring the motion equation (\ref{nd:CHeqn}) to preserve
${\rm div}\,\mathbf{u}=0={\rm div}\,\mathbf{v}$ implies a Poisson
equation for the pressure $p$ with a Neumann boundary condition, which
is obtained just as usual in the case of incompressible ideal fluid
dynamics, by taking the normal component of the motion equation
evaluated at the boundary. Of course, the $n$-dimensional extension
of the CH equation (\ref{nd:CHeqn}) reduces to Euler's equation when we
set $\alpha = 0$. The properties of the $n$-dimensional
CH equation (\ref{nd:CHeqn}) are summarized in Holm, Marsden and Ratiu
[1998a,b] and Holm, Kouranbaeva, Marsden, Ratiu, and Shkoller [1998]
for the ideal case. In particular, equation (\ref{nd:CHeqn}) is shown
to be the geodesic spray equation for geodesic motion on the group
$\mbox {\rm Diff}(\mathcal{D})$ with respect to the metric given by
the $H^1$ norm of the fluid velocity
$\mathbf{u}$ in $n$ dimensions. See also Chen, Foias, Holm, Olson,
Titi and Wynne [1998] and  Foias, Holm and Titi [1998] for discussions
of the corresponding viscous, forced case and its connection to mean
turbulence closure models.

\paragraph{Discussion of the CH Equation.}
The essential idea of the CH equation is that its specific momentum
(i.e., its momentum per unit mass) is transported by a velocity
which is smoothed by inverting the elliptic Helmholtz operator
$(1-\alpha^2\Delta)$, where $\alpha$ corresponds to the length scale
at which this smoothing becomes important, i.e., when it becomes of
order $O(1)$. When the smoothing operator $(1-\alpha^2\Delta)^{-1}$
is applied to the transport velocity in Euler's equation to produce
the CH equation, its effect on length scales smaller than $\alpha$
is that steep gradients of the specific momentum $\mathbf{v}$ tend not
to steepen much further, and thin vortex tubes tend not to get much
thinner as they are transported. And, its effect on length scales
that are considerably larger than $\alpha$ is negligible. Hence, the
transport of vorticity in the CH equation is intermediate between
that for the Euler equations in two and three dimensions. As for
Euler vorticity, the curl of the CH specific momentum is {\em
convected} as an {\em active} two form, but its transport velocity
is the {\em smoothed, or filtered} CH specific momentum.

The effects of this smoothing or filtering of the transport velocity
in the CH equation can be seen quite clearly from its Fourier
spectral representation in the periodic case. In this case, we define
$\mathbf{v}_\mathbf{k}$ as the $\mathbf{k}$-th Fourier mode of the
specific momentum $\mathbf{v}\equiv(1-\alpha^2\Delta)\mathbf{u}$ for
the CH equation; so that $\mathbf{v}_\mathbf{k} \equiv (1+\alpha^2
|\mathbf{k}|^2)\mathbf{u}_\mathbf{k}$. Then the Fourier spectral
representation of the CH equation for a periodic three-dimensional
domain is expressed as
\begin{equation} \label{CH-spectral}
\Pi\left(
\frac{d}{dt} \mathbf{v}_\mathbf{k}\,
- i \sum_{\mathbf{p}+\mathbf{n}=\mathbf{k}}
\frac{ \mathbf{v}_\mathbf{p} }{1+\alpha^2|\mathbf{p}|^2 }
\times ( \mathbf{n}\times\mathbf{v}_\mathbf{n}) \right)
=0,
\end{equation}
where $\Pi$ is the Leray projection onto Fourier modes transverse to
$\mathbf{k}$. (As usual, the Leray projection ensures
incompressibility.) In this Fourier spectral representation of the
CH equation, one sees that the coupling to high modes is suppressed
by the denominator when $1 + \alpha^2|\mathbf{p}|^2\gg1$.
Consequently, when $|\mathbf{p}|\geq{O(1/\alpha)}$, the smoothing of
the transport velocity suppresses the Fourier-mode interaction
coefficients. In fact, the CH smoothing of the transport velocity
suppresses {\em both} the forward and backward cascades for wave
numbers $|\mathbf{p}|\geq{O(1/\alpha)}$, but leaves the Euler dynamics
essentially unchanged for smaller wave numbers. The
result is that the vortex stretching term in the dynamics of
$\mathbf{q}={\rm curl}\,\mathbf{v}$ is mollified in the CH model and
so the vortices at high wave numbers will tend to be ``shorter and
fatter'' than in the corresponding Euler case for the same initial
conditions.

\paragraph{The Kelvin--Noether Theorem.} Since the $n$-dimensional CH
equation (\ref{nd:CHeqn}) is Euler--Poincar\'e, it also has a
corresponding {\bfi Kelvin--Noether circulation theorem}. Namely,
\begin{equation}
\frac{ dI}{dt}=0
\quad \hbox{where} \quad
I(t)=
\oint_{\gamma_t}(\mathbf{u} - \alpha^2 \Delta \mathbf{u})\cdot
d\mathbf{x} =
\oint_{\gamma_t}\mathbf{v}\cdot
d\mathbf{x}\,,
\label{KelThm2}
\end{equation}
for any closed curve ${\gamma_t}$ that moves with the fluid velocity
$\mathbf{u}$. This expression for the Kelvin--Noether circulation of
the CH equation in three dimensions is reminiscent of the
corresponding expression (\ref{I-psmom}) in GLM theory involving the
pseudomomentum for wave, mean-flow interaction theory. This
correspondence confirms the physical interpretation of the $\alpha^2$
term in the Kelvin--Noether circulation integral as a {\it Lagrangian
mean closure relation} for the pseudomomentum of the high frequency
(i.e., rapidly fluctuating, turbulent) components of the flow. In this
interpretation, $\alpha$ for the CH equation corresponds to both the
amplitude of these high frequency components and the length scale at
which they become important.

\paragraph{Energy Conservation for the CH Equation.} Legendre
transforming the action (\ref{CH-lag}) gives the following conserved
{\bfi Hamiltonian} (still expressed in terms of the velocity, instead
of the momentum density
$\mathbf{m}= \delta l/ \delta \mathbf{u}$),
\begin{equation}
H = \int_{\cal M} d^{\,n}x\ \Big[\,\frac{D}{2}
\left(|\mathbf{u}|^2+\alpha^2|\nabla\mathbf{u}|^2\right)
 + p(D-1)\Big]
\,.  \label{CH-ham-u}
\end{equation}
Thus, when evaluated on the constraint manifold $D=1$, the Lagrangian
and the Hamiltonian for the CH equation coincide in $n$ dimensions.
(This, of course, is not unexpected for a stationary principle giving
rise to geodesic motion.)

The curl of the 3D Camassa-Holm motion equation (\ref{nd:CHeqn}) yields
\begin{equation} \label{vortex-stretching}
\frac{ \partial}{\partial t}\mathbf{q}
= \mathbf{q}\cdot\nabla\mathbf{u} - \mathbf{u}\cdot\nabla\mathbf{q}
\equiv [\mathbf{u},\mathbf{q}\,],
\quad \hbox{where} \quad
\mathbf{q}\equiv{\rm curl}
(\mathbf{u} - \alpha^2 \Delta \mathbf{u})\,,
\end{equation}
and we have used incompressibility and commutativity of the
divergence and Laplacian operators. Thus, $\mathbf{u}$ is the
transport velocity for the generalized vorticity $\mathbf{q}$ and the
``vortex stretching" term  $\mathbf{q}\cdot\nabla\mathbf{u}$ involves
$\nabla\mathbf{u}$, whose $L^2$ norm is {\em controlled} by the
conservation of energy in equation (\ref{CH-ham-u}). Boundedness of
this norm will be useful in future analytical studies of the 3D
Camassa-Holm equation; for example, in the investigation of the
Liapunov stability properties of its equilibrium solutions.

\paragraph{The Riemannian CH Equation.} One can formulate the CH
equation on a general Riemannian manifold, possibly with boundary.
Although this formulation will be the subject of future papers, we
comment on it here because of the importance of spherical geometry,
in particular, for GFD models.

\subsection{The Euler-Boussinesq $\alpha$ Model (EB$\alpha$)} We
introduce nonlinear dispersion into the Euler-Boussinesq equations
using the Euler--Poincar\'e framework and following the example of the
CH equations.

\paragraph{The Lagrangian.} To carry this out, we modify the EB
Lagrangian (\ref{lag-v1}) by simply adding the $\alpha^2$ term (while
dropping other scale factors
$\epsilon$, $\sigma$, as well as primes and subscripts)
\begin{eqnarray}\label{MEB-lag}
\mathfrak{S}_{{\rm EB}\alpha}
\!\!\!\!&=&\!\!\!\! \int dt \; l_{{\rm EB}\alpha}
\\
\!\!\!\!&=&\!\!\!\! \int\!\! dt\!\!\int_{\cal M}\!\!
d^{\,n}x \Big[
\,D \Big(\frac{1}{2}|\mathbf{u}|^2
+ \frac{\alpha^2}{2} |\nabla\mathbf{u}|^2
+ \mathbf{u} \cdot \mathbf{R}( \mathbf{x} ) - g b z
\Big) - p(D-1)\Big]\,,
\nonumber
\end{eqnarray}
This action is the order $O(\alpha^2)$ approximation of the mean of the
EB action in equation (\ref{l}), using the Taylor expansion
(\ref{E-avg}) and neglecting order $O(\alpha^2\epsilon)$ pressure and
density corrections. (An order $O(\alpha^2)$ term involving
$\overline{({\boldsymbol{\xi}}\cdot \nabla \mathbf{u})
({\boldsymbol{\xi}}\cdot \nabla \mathbf{R})}$ is also dropped, since it
makes only a negligible contribution in the resulting motion equation.)

\paragraph{The Euler--Poincar\'e Equations.} Varying this action at
fixed
$\mathbf{x}$ and
$t$ gives
\begin{eqnarray}
\delta \mathfrak{S}_{{\rm EB}\alpha}
&=&\!\! \int dt\int_{\cal M} d^{\,n}x \,\Big[
\left( \frac{1}{2}|\mathbf{u}|^2
+ \frac{\alpha^2}{2}|\nabla\mathbf{u}|^2
+ \mathbf{u} \cdot \mathbf{R}( \mathbf{x} )
- g b z-p\right) \delta D
\nonumber \\
&&
 - {Dgz \delta b} - (D-1)\delta p
+ \left(D \mathbf{u}
- \alpha^2 ({\rm div}D{\rm grad}) \mathbf{u} \right)\cdot
\delta \mathbf{u}\,\Big]
\nonumber \\
&&\hspace{.5in}
+\ \alpha^2 \int dt\oint_{\partial{\cal M}} d^{\,n-1}x \
(D\hat{\mbox{\boldmath{$n$}}}\cdot \nabla \mathbf{u} \cdot \delta \mathbf{u})
\,,
\label{var.MEB}
\end{eqnarray}
where the natural boundary conditions are again given in
(\ref{boundary.equation}).
The corresponding Euler--Poincar\'e equation for
the action $\mathfrak{S}_{{\rm EB}\alpha}$
in equation (\ref{MEB-lag}) is,
\begin{eqnarray}
&&\left(\frac{ \partial}{\partial t}
+ \mathbf{u}\cdot\nabla\right)\mathbf{v}
- \mathbf{u}\times {\rm curl} \mathbf{R}
+ v_j\nabla u^j
+ gb \hat{\mbox{\boldmath{$z$}}}
\nonumber \\
&&\hspace{1in}
+\nabla\left(p - \frac{1}{2}|\mathbf{u}|^2
- \frac{\alpha^2}{2}|\nabla\mathbf{u}|^2\right) = 0\,,
\label{MEB-eqn}
\end{eqnarray}
where
\begin{eqnarray} \label{def-MEB}
&&\mathbf{v} \equiv \mathbf{u} - \alpha^2\Delta \mathbf{u}, \quad
\nabla \cdot \mathbf{u} = 0,
\nonumber \\
&&\frac{db}{dt} = 0, \quad \frac{d}{dt}
= \left(\frac{\partial}{\partial t}
+ \mathbf{u} \cdot \nabla \right)\,.
\end{eqnarray}
Again, we have substituted the constraint $D=1$, which implies
incompressibility via the continuity equation for $D$. Relative to
the usual EB equations (\ref{EB-mot}), the EB$\alpha$ equation
(\ref{MEB-eqn}) has a smoothed, or filtered transport velocity,
since $\mathbf{u} = (1-\alpha^2\Delta)^{-1}\mathbf{v}$.

\paragraph{The Kelvin--Noether Theorem.} The Kelvin--Noether
circulation theorem for the EB$\alpha$ equation (\ref{MEB-eqn}) is,
\begin{equation}\label{KN-theorem-MEB}
\frac{d}{dt}\oint_{\gamma_t(\mathbf{u})}
(\mathbf{v}+\mathbf{R})\cdot d\mathbf{x}
= -\oint_{\gamma_t(\mathbf{u})} b dz\;,
\end{equation}
where the curve $\gamma_t(\mathbf{u})$ moves with the fluid velocity
$\mathbf{u}$. (The two Kelvin theorems in equations
(\ref{KN-theorem-MEB}) and (\ref{KN-theorem-EB})  differ in their
definitions of $\mathbf{v}$.) By Stokes' theorem, the EB$\alpha$
equations generate circulation of  $\mathbf{v}+\mathbf{R}$ around
$\gamma_t(\mathbf{u})$ whenever the gradient of bouyancy is not
vertical. The conservation of potential vorticity on fluid parcels
for the EB$\alpha$ equations is given by
\begin{equation} \label{pv-cons-MEB}
\frac{\partial q_{{\rm EB}\alpha}}{\partial{t}}
+\mathbf{u}\cdot\nabla\,q_{{\rm EB}\alpha} = 0\,,
\quad \hbox{where}\quad
q_{{\rm EB}\alpha}=\nabla{b}\cdot\nabla\times(\mathbf{v}+\mathbf{R})\,.
\end{equation}
The curl of the EB$\alpha$ motion equation gives,
\begin{equation} \label{vortex-stretching-MEB}
\frac{ \partial}{\partial t}\mathbf{q}
= - \mathbf{u}\cdot\nabla\mathbf{q}
+ \mathbf{q}\cdot\nabla\mathbf{u}
+ g \nabla b\times \hat{\mbox{\boldmath{$z$}}},
\quad \hbox{where} \quad
\mathbf{q}\equiv{\rm curl}
(\mathbf{u} - \alpha^2 \Delta \mathbf{u} + \mathbf{R})\,.
\end{equation}
This the usual expression for transport, stretching and creation of
vorticity in a buoyant flow, except that here the vortex stretching
coefficient is $\nabla\mathbf{u}$, which is {\it moderated} relative to
the usual EB equations.

\paragraph{Energy Conservation.} The EB$\alpha$ equations
(\ref{MEB-eqn}) conserve the following Hamiltonian (found, e.g., by
Legendre transforming the Lagrangian
$l_{{\rm EB}\alpha}$ in equation (\ref{MEB-lag})). Namely,
\begin{equation} \label{MEB-Ham}
H_{{\rm EB}\alpha} = \int_{\cal M}
d^{\,n}x \Big[
\,D \Big(\frac{1}{2}|\mathbf{u}|^2
+ \frac{\alpha^2}{2} |\nabla\mathbf{u}|^2\Big)
+  Dgbz
+ p(D-1)\Big]\,.
\end{equation}
The corresponding conserved energy is
\begin{equation} \label{MEB-erg}
E_{{\rm EB}\alpha} = H_{{\rm EB}\alpha}\Big|_{D=1} = \int_{\cal M}
d^{\,n}x \Big[
\frac{1}{2}|\mathbf{u}|^2 + \frac{\alpha^2}{2} |\nabla\mathbf{u}|^2
+  gbz
\Big]\,.
\end{equation}
Since the (finite) value of this conserved energy for the EB$\alpha$
model is determined by its initial conditions and $b$ is advected (so
it has a maximum value in $L^{\infty}$) there is $H^1$ control of the
velocity $\mathbf{u}$. This is the effect of the ``filtering'' of the
solution produced by the nonlinear dispersion for $\alpha\ne0$. This
filtering moderates the growth of instabilities at wavenumbers
$|{\bf k}|\ge 1/ \alpha $. So if the EB$\alpha$ model is used as a
large eddy simulation (LES) model, one would choose the value of
$ \alpha $ to determine the size of the minimum resolved length
scale. The filtering by the $\alpha$ term also allows nonlinear
Liapunov stability conditions to be formulated for equilibrium
solutions of the EB$\alpha$ model. This stability result is clear from
the work of Abarbanel et al. [1986], who introduced the notion of
``conditional'' Liapunov stability for the EB model, using wavenumber
conditions that now turn out to be satisfied for the EB$\alpha$ model.
The Euler fluid equations (\ref{Eul-mot-nd}) may also be modified
analogously to include nonlinear dispersion. However, this case is
ignored for now, as we proceed to discuss the modified primitive
equations.

\subsection{Primitive Equation $\alpha$ Model (PE$\alpha$)}
In horizontal and vertical components, with
$\mathbf{v}^{\,\perp}\equiv\mathbf{v}
- v^{\|}\hat{\mbox{\boldmath{$z$}}}$,
$v^{\|}\equiv(\mathbf{v}\cdot \hat{\mbox{\boldmath{$z$}}})$ and
${\rm curl}\,\mathbf{R}=f(\mathbf{x})\hat{\mbox{\boldmath{$z$}}}$,
the the EB$\alpha$ equations are expressed in nondimensional
form as, cf. equations
(\ref{h+v.EB}) and  (\ref{MEB-eqn}),
\begin{eqnarray} \label{hv-MEB-eqn}
\epsilon\,\frac{ d\mathbf{v}^{\,\perp} }{dt}
+ \epsilon\,v_j \nabla^{\,\perp} u^j
+ f\hat{\mbox{\boldmath{$z$}}}\times \mathbf{u}
+ \nabla^{\,\perp}\, \pi = 0\,,
&&
\epsilon\, \sigma^2\,\frac{dv^{\|}}{dt}+ b +
\frac{\partial{\pi}}{\partial{z}}   = 0\,,
\\
\frac{d}{dt} \equiv
\left(\frac{\partial}{\partial t}
+ \mathbf{u}^{\,\perp} \cdot \nabla^{\,\perp}
+ {u^{\|}}\frac{\partial}{\partial z} \right),
\quad \frac{db}{dt}=0\,,
&&
\nabla\cdot\mathbf{u}=\nabla^{\,\perp}\cdot\mathbf{u}^{\,\perp}
+\frac{\partial{u^{\|}}}{\partial{z}}=0\,,
\nonumber \\
\hbox{where} \quad
\pi \equiv \left(p - \frac{1}{2}|\mathbf{u}|^2
- \frac{\alpha^2}{2}|\nabla\mathbf{u}|^2\right),
&&
\mathbf{v} \equiv \mathbf{u} - \alpha^2\Delta \mathbf{u}\,.
\nonumber
\end{eqnarray}
Here, $\epsilon$ and $\sigma$ are the Rossby number and aspect ratio,
respectively, and $\alpha$ has been scaled in units of horizontal
length scale, $L$. The leading order balances are still hydrostatic
in the vertical, and geostrophic in the horizontal. Setting $ \sigma
=0$ in equation (\ref{hv-MEB-eqn}) removes the vertical acceleration
and thereby produces the primitive equation $\alpha$ model
(PE$\alpha$). We expect that the filtering property for
$\alpha\ne0$ discussed above for the CH and EB$\alpha$ equations should
make the PE$\alpha$ model much more regular than the ordinary PE, from
the viewpoint of gravity wave oscillations. The problematic asspects
of gravity waves in atmospheric and oceanic numerical simulations and
data assimilation using the PE model have often been addressed by
invoking the concept of an idealized ``slow manifold'' on which
gravity waves are absent, see, e.g., Lorenz [1992]. However, the
existence of a slow manifold has never been proven for the PE model. It
is an open question, whether the new PE$\alpha$ model will have a slow
manifold when dissipation and forcing are included. We shall report on
this matter elsewhere.

\subsection*{Final Remarks}
\addcontentsline{toc}{section}{Final Remarks}
In this paper we have shown how asymptotic expansions in Hamilton's
principle for the Euler--Poincar\'e equations of geophysical fluid
dynamics provide an organizing principle for many GFD systems and
produce a clear, unified understanding of their Kelvin theorems.
Hamilton's principle asymptotics in the Euler--Poincar\'e setting thus
explains the shared properties of these GFD models and provides a
unified approach to making additional approximations and creating new
models. In this setting, we have introduced a new class of fluid
models, called $\alpha$ models, that possess nonlinear dispersion which
smooths the transport velocity relative to the circulation velocity in
the Euler--Poincar\'e equations. The effect of this smoothing is to
moderate the vortex stretching process while preserving the Kelvin
circulation theorem for these equations. The efficacy and utility of
the $\alpha$ models are yet to be determined, but initial studies of
them are promising. We expect that one can also perform the sort of
asymptotics done here on the group level first (to get an
approximating group) and then perform Euler--Poincar\'e reduction and
arrive at the same conclusions. In other words, there should be a
general principle for asymptotics in Hamilton's principle, which
allows passage from one group and its corresponding advected
quantities to an approximating one. Moreover, this process should
commute with Lagrangian and Hamiltonian reduction.

\subsection*{Acknowledgments}
\addcontentsline{toc}{section}{Acknowledgments}
We would like to extend our gratitude to John Allen, Ciprian Foias,
Rodney Kinney, Hans-Peter Kruse, Len Margolin, Jim McWilliams, Balu
Nadiga, Ian Roulstone, Steve Shkoller, Edriss Titi and Vladimir
Zeitlin for their time, encouragement and invaluable input. Work by
D. Holm was conducted under the auspices of the US Department of
Energy, supported (in part) by funds provided by the University of
California for the conduct of discretionary research by Los Alamos
National Laboratory. Work of J. Marsden was supported by the
California Institute of Technology and NSF grant DMS
96--33161. Work by T. Ratiu was partially supported by NSF Grant
DMS-9503273  and DOE contract DE-FG03-95ER25245-A000. Some of this
research was performed while two of the authors (DDH and TSR) were
visiting the Isaac Newton Institute for Mathematical Sciences at
Cambridge University. We gratefully acknowledge financial support by
the program ``Mathematics in Atmosphere and Ocean Dynamics" as well
as the stimulating atmosphere at the Isaac Newton Institute during our
stay there.

\subsection*{References}
\addcontentsline{toc}{section}{References}
\begin{description}

\item
Abarbanel, H.D.I., and Holm, D.D. 1987,
Nonlinear stability of inviscid flows in three
dimensions: incompressible fluids and barotropic fluids.
{\it Phys. Fluids} {\bf 30} (1987), 3369--3382.

\item
Abarbanel, H.D.I., Holm, D.D., Marsden, J.E. and Ratiu, T. 1986,
Nonlinear stability analysis of stratified ideal fluid equilibria.
{\it Phil Trans. Roy. Soc.} (London) A {\bf 318}, 349--409.

\item Allen, J.S. 1991,
Balance equations based on momentum equations with global invariants
of potential enstrophy and energy.
{\it J. Phys. Oceanogr.} {\bf 21}, 265--276.

\item Allen, J.~S., Barth, J.~A., and Newberger, P.~A.
1990a, On intermediate models for barotropic continental shelf and
slope flow fields. Part I: Formulation and comparison of exact
solutions. {\it J. Phys. Oceanogr.} {\bf 20}, 1017--1042.

\item Allen, J.~S., Barth, J.~A., and Newberger, P.~A.
1990b, On intermediate models for barotropic continental shelf and
slope flow fields. Part III: Comparison of numerical model
solutions in periodic channels. {\it J. Phys. Oceanogr.} {\bf 20},
1949--1973.

\item Allen, J.~S. and Holm, D.D. 1996,
Extended-geostrophic Hamiltonian models for rotating
shallow water motion.
{\it Physica D}, {\bf 98} (1996) 229--248.

\item Allen, J.~S., Holm, D.D. and Newberger, P. 1998,
Toward an extended-geostrophic model Euler--Poincar\'e
model for mesoscale oceanographic flow.
{\it These Proceedings}.

\item Allen, J.~S., and Newberger, P.~A. 1993,
On intermediate models for stratified flow.
{\it J. Phys. Oceanogr.} {\bf 23}, 2462--2486.

\item Andrews, D.G., and McIntyre, M.E. 1978a,
An exact theory of nonlinear waves on a Lagrangian-mean flow.
{\it J. Fluid Mech.} {\bf 89}, 609--646.

\item Andrews, D.G., and McIntyre, M.E. 1978b,
On wave-action and its relatives.
{\it J. Fluid Mech.} {\bf 89}, 647--664, addendum {\it ibid} {\bf 95},
796.

\item Arnold, V.I., 1966,
Sur la g\'{e}om\'{e}trie differentielle
des groupes de Lie de dimenson
infinie et ses applications \`{a}
l'hydrodynamique des fluids parfaits.
{\it Ann. Inst. Fourier, Grenoble\/} {\bf 16}, 319--361.

\item Barth, J.~A., Allen, J.~S., and Newberger,
P.~A., 1990, On intermediate models for barotropic continental shelf
and slope flow fields. Part II: Comparison of numerical model
solutions in doubly periodic domains. {\it J. Phys. Oceanogr.} {\bf
20}, 1044--1076.

\item Bloch, A.M., Krishnaprasad, P.S.,
Marsden, J.E., and Murray, R. 1996,
Nonholonomic mechanical systems with symmetry.
{\it Arch. Rat. Mech. An.}, {\bf 136}, 21--99.

\item Bloch, A.M., Krishnaprasad, P.S.,
Marsden, J.E. and Ratiu, T.S. 1996,
The Euler-Poincar\'{e} equations and
double bracket dissipation.
{\it Comm. Math. Phys.}  {\bf 175}, 1--42.

\item Bretherton, F.P., 1970
A note on Hamilton's principle for perfect fluids.
{\it J. Fluid Mech.\/} {\bf
44}, 19--31.

\item Browning, G.L., Holland, W.R., Kreiss, H.-O. and
Worley, S. J. 1990, An accurate hyperbolic system for
approximately hydrostatic and incompressible oceanographic flows.
{\it Dyn. Atm. Oceans} {\bf 14} 303--332.

\item Camassa, R., and Holm, D.D., 1993,
An integrable shallow water equation
with peaked solitons.
{\it Phys. Rev. Lett.} {\bf 71} (1993) 1661--1664.

\item Camassa, R., Holm, D.D. and Levermore, C.D. 1995,
Long-time shallow water equations with a varying bottom.
{\it Physica D}, {\bf 98} (1996) 258--286.

\item  Cendra, H., Holm, D.D.,
Hoyle, M.J.W. and Marsden, J.E. 1997,
The Maxwell-Vlasov equations in Euler-Poincar\'{e} form.
{\it J. Math. Phys.}, to appear

\item  Cendra, H., Holm, D.D., Marsden, J.E. and Ratiu, T.S. 1997,
Lagrangian reduction, the Euler--Poincar\'{e} equations, and semidirect
products. {\it Arnold volume II, Am. Math. Soc.}, to appear.

\item Cendra, H., Ibort, A. and Marsden, J.E. 1987,
Variational principal fiber bundles: a geometric
theory of Clebsch potentials
and Lin constraints. {\it J.  Geom.  Phys.\/} {\bf 4}, 183--206.

\item Cendra, H. and Marsden, J.E. 1987,
Lin constraints, Clebsch potentials and variational principles.
{\it Physica D\/} {\bf 27}, 63--89.

\item  Cendra, H., Marsden, J.E. and Ratiu, T.S. 1997,
Lagrangian reduction by stages. {\it Preprint.}

\item Charney, J.G., 1948,
On the scale of atmospheric motions.
{\it J. Meteorol.}, {\bf 4}, 135--163.

\item Chen, S.Y., Foias, C., Holm, D.D., Olson, E., Titi, E.S.
and Wynne, S. 1998, The steady viscous Camassa--Holm equations as a
Reynolds algebraic stress closure model for turbulent channel flows
and their boundary layer analysis. {\it In preparation}.

\item Craik, A.D.D., and Leibovich, S. 1976, A rational model
for Langmuir circulations. {\it J. Fluid Mech.} {\bf 73}, 401--426.

\item Cullen, M.J.P., and Purser, R.J. 1989,
Properties of the Lagrangian semigeostrophic equations.
{\it J. Atmos. Sci.} {\bf46}, 1477--1497.

\item Cushman-Roisin, B., 1994,
{\it Introduction to Geophysical Fluid Dynamics}.
Prentice-Hall, New Jersey.

\item Eliassen, A., 1949,
The quasi-static equations of motion with pressure as
independent variable.
{\it Geofysiske Publikasjoner} {\bf 17} (3) 1--44.
Utgitt av Det Norske Videnskaps-Akademi, Oslo.

\item Foias, C., Holm, D.D., and Titi, E.S. 1997,
Global well-posedness for the three dimensional
viscous Camassa--Holm equation. {\it In preparation}.

\item Gent, P.R., and McWilliams, J.C. 1983a,
Consistent balanced models in bounded and periodic domains.
{\it Dyn. Atmos. Oceans} {\bf 7}, 67--93.

\item Gent, P.R., and McWilliams, J.C. 1983b,
Regimes of validity for balanced models.
{\it Dyn. Atmos. Oceans} {\bf 7}, 167--183.

\item Gent, P.R., and McWilliams, J.C. 1984,
Balanced models in isentropic coordinates and the shallow water
equations.
{\it Tellus} {\bf 36A}, 166--171.

\item Gent, P.R., McWilliams, J.C., and
Snyder, C. 1994,
Scaling analysis of curved fronts: validity of the balance
equations and semigeostrophy.
{\it J. Atmos. Sci.} {\bf 51}, 160--163.

\item  Gjaja, I. and and Holm, D.D., 1996,
Self-consistent wave-mean flow interaction
dynamics and its Hamiltonian formulation for a rotating
stratified incompressible fluid.
{\it Physica D}, {\bf 98} (1996) 343--378.

\item Holm, D.D. 1996,
Hamiltonian balance equations.
{\it Physica D}, {\bf 98} 379--414

\item Holm, D.D., Kouranbaeva, S., Marsden, J.E.,
Ratiu, T. and Shkoller, S. 1998,
Euler-Poincar\'{e} equations for continuum theories
on Riemannian manifolds. {\it In preparation}.

\item Holm, D.D., and Kupershmidt, B.A. 1983,
Poisson brackets and Clebsch representations
for magnetohydrodynamics, multifluid  plasmas, and elasticity.
{\it Physica D} {\bf 6} (1983), 347--363.

\item Holm, D.D., Kupershmidt, B.A., and Levermore, C.D. 1983,
Canonical maps between Poisson brackets in
Eulerian and Langrangian descriptions of continuum mechanics.
{\it Phys. Lett. A} {\bf 98}, 389--395.

\item Holm, D.D., Lifschitz, A. and Norbury, J. 1998,
Eulerian semigeostrophic shallow water dynamics.
{\it In preparation}.

\item Holm, D.D. and Long, B. 1989,
Lyapunov stability of ideal
stratified fluid equilibria in hydrostatic balance.
{\it Nonlinearity} {\bf 2}, 23--35.

\item Holm, D.D., Marsden, J.E., and Ratiu, T. 1987,
{\it Hamiltonian Structure and Lyapunov Stability for Ideal
Continuum Dynamics}. ISBN 2-7606-0771-2, University of Montreal
Press: Montreal.

\item Holm, D. D., Marsden, J. E. and Ratiu, T. 1998a,
The Euler-Poincar\'{e} equations and semidirect products with
applications to continuum theories. {\it Adv. in Math.}, to appear.

\item Holm, D. D., Marsden, J. E. and Ratiu, T. 1998b,
Euler--Poincar\'e models of ideal fluids
with nonlinear dispersion. {\it Phys. Rev. Lett.}, to appear.

\item Holm, D.D., Marsden, J.E., Ratiu, T., and Weinstein, A. 1985,
Nonlinear stability of fluid and plasma equilibria.
{\it Physics Reports}  {\bf 123}, 1--116.

\item Holm, D.D., and Zeitlin, V. 1997,
Hamilton's principle for quasigeostrophic motion.
{\it Phys. Fluids A}, to appear.

\item Hoskins, B.J., 1975,
The geostrophic momentum approximation and the
semigeostrophic equations.
{\it J. Atmos. Sci.} {\bf32}, 233--242.

\item Lorenz, E.N. 1960,
Energy and numerical weather prediction.
{\it Tellus} {\bf 12}, 364--373.

\item Lorenz, E.N. 1992,
The slow manifold: what is it?
{\it J. Atm. Sci.} {\bf 49}, 2449-2451.

\item Marsden, J.E., Patrick, G.W. and Shkoller, S. 1997
Multisymplectic geometry, variational integrators,
and nonlinear PDEs. {\it Preprint}.

\item Marsden, J.E. and Ratiu, T.S. 1994,  {\it Introduction to
Mechanics and Symmetry.\/} Texts in Applied Mathematics, {\bf  17},
Springer-Verlag.

\item Marsden, J.E. and Scheurle, J. 1993a,
Lagrangian reduction and the double spherical pendulum.
{\it ZAMP\/} {\bf 44}, 17--43.

\item Marsden, J.E. and Scheurle, J. 1993b,
The reduced Euler-Lagrange equations.
{\it Fields Institute Comm.\/} {\bf 1}, 139--164.

\item Marsden, J.E. and Weinstein, A. 1982,
The Hamiltonian structure of the Maxwell-Vlasov equations.
{\it Physica D\/} {\bf 4}, 394--406.

\item Marsden, J.E., Weinstein, A., Ratiu, T.S.
Schmid, R. and Spencer, R.G. 1983,
Hamiltonian systems with symmetry, coadjoint
orbits and plasma physics. In
Proc. IUTAM-IS1MM Symposium on
{\it Modern Developments in Analytical Mechanics\/},
Torino 1982, {\it Atti della Acad. della Sc. di Torino\/}
{\bf 117}, 289--340.

\item McWilliams, J.C., and Gent, P.R. 1980,
Intermediate models of planetary circulations in the
atmosphere and ocean.
{\it J. Atmos. Sci.} {\bf 37}, 1657--1678.

\item McWilliams, J.C., and Gent, P.R. 1986,
The evolution of sub-mesoscale coherent vortices on the
$\beta$-plane.
{\it Geophys. Astrophys. Fluid Dyn.} {\bf 35}, 235--255.

\item McWilliams, J.C., Gent, P.R., and Norton, N.J. 1986,
The evolution of balanced, low-mode vortices  on the $\beta$-plane.
{\it J. Phys. Oceanogr.} {\bf 16}, 838--855.

\item McWilliams, J.C., Norton, N.J., Gent, P.R., and
Haidvogel, D.B. 1990,
A linear balance model of wind-driven, midlatitude ocean circulation.
{\it J. Phys. Oceanogr.} {\bf 20}, 1349--1378.

\item Norton, N.J., McWilliams, J.C., and Gent, P.R.
1986, A numerical model of the balance equations in a periodic
domain and an example of balanced turbulence.
{\it J. Comput. Phys.} {\bf 67}, 439--471.

\item Pedlosky, J. 1987, {\it Geophysical Fluid Dynamics}, 2nd
Edition, Springer, New York.

\item Phillips, N. A., 1963,
Geostrophic motion.
{\it Rev. Geophys.}, {\bf1}, 123--126.

\item Phillips, O.M., 1969, {\it The Dynamics of the Upper Ocean},
Cambridge University Press, Cambridge.

\item Poincar\'{e}, H. 1885,
Sur l'\'{e}quilibre d'une masse fluide anim\'{e}e d'un mouvement de
rotation.  {\it Acta. Math.\/} {\bf 7}, 259.

\item Poincar\'{e}, H. 1890, {\it Th\'eorie des tourbillons},
Reprinted by \'Editions Jacques Gabay, Paris.

\item Poincar\'{e}, H. 1890,
Sur le probl\`{e}me des trois corps et les \'{e}quations de la
dynamique. {\it Acta Math.\/} {\bf 13}, 1--271.

\item Poincar\'e, H. 1892--1899,
{\it Les M\'ethodes Nouvelles de la M\'ecanique Celeste.\/}
3 volumes. English translation {\it New Methods of
Celestial Mechanics.\/} History of Modern Physics and
Astronomy {\bf 13}, Amer. Inst. Phys., 1993.

\item Poincar\'{e}, H. 1892, Les formes d'\'{e}quilibre d'une
masse fluide en rotation.  {\it Revue G\'{e}n\'{e}rale des
Sciences\/} {\bf 3}, 809--815.

\item Poincar\'{e}, H. 1901a, Sur la stabilit\'{e} de
l'\'{e}quilibre  des figures piriformes affect\'{e}es par une
masse fluide en rotation. {\it Philosophical Transactions A\/}
{\bf 198}, 333--373.

\item Poincar\'{e}, H. 1901b, Sur une forme nouvelle des
\'{e}quations de la m\'{e}chanique.  {\it C.R. Acad. Sci.\/} {\bf
132}, 369--371.

\item Poincar\'{e}, H. 1901,
Sur une forme nouvelle des \'{e}quations de la m\'{e}chanique.
{\it CR Acad. Sci.\/} {\bf 132}, 369--371.

\item Poincar\'{e}, H. 1910, Sur la precession des corps
deformables. {\it Bull Astron\/} {\bf 27}, 321--356.

\item Rossby, C.G., 1940,
Planetary flow patterns in the atmosphere.
{\it Quart. J. R. Met. Soc.} {\bf 66}, Suppl., 68--87.

\item Roulstone, I., and Brice, S.J. 1995,
On the Hamiltonian formulation of the quasi-hydrostatic
equations.
{\it Q. J. R. Meteorol. Soc} {\bf 121}, 927--936.

\item Salmon, R., 1983,
Practical use of Hamilton's principle.
{\it J. Fluid Mech.} {\bf 132}, 431--444.

\item Salmon, R., 1985,
New equations for nearly geostrophic flow.
{\it J. Fluid Mech.} {\bf 153}, 461--477.

\item Salmon, R., 1988,
Hamiltonian fluid dynamics.
{\it Ann. Rev. Fluid Mech.} {\bf 20}, 225--256.

\item Weinstein, A., 1983, Hamiltonian structure for drift waves
and geostrophic flow. {\it Phys. Fluids} {\bf 26}, 388--390.

\item White, A.A., 1977, Modified quasi-geostrophic equations
using height as a vertical coordinate.
{\it Q. J. R. Metereol. Soc.} {\bf 103}, 383--396.

\item Xu, Q. 1994
Semibalance model --- connection between geostrophic-type and
balanced-type intermediate models.
{\it J. Atmos. Sci.} {\bf 51}, 953--970.

\end{description}

\end{document}